# Scale-Separated Collective Bubble Nucleation and Departure

Rameez Iqbal,[1] Gautier Rouaze,[1] Gio Bellone,[1] Francesco Minola,[1] Lenan Zhang,[2] Zhengmao Lu[1*]

[1]Energy Transport Advances Laboratory (ETA-Lab), Institute of Mechanical Engineering, EPFL, 1015, Lausanne, Switzerland

[2]Energy Research Lab, Sibley School of Mechanical and Aerospace Engineering, Cornell University, Ithaca, 14853, USA

*Corresponding author: zhengmao.lu@epfl.ch

Nucleation is classically treated as a local process, yet whether coupling between neighboring sites governs activation and stability remains unexplored. Here we show bubble nucleation is fundamentally collective: sites separated by the hydrodynamic-boundary-layer scale activate more readily and resist deactivation under changing thermal loads, consistent with a non-local hydrodynamic shielding mechanism, whereby neighboring bubbles slow the intervening flow, suppress convective heat removal, and stabilize vapor embryos. Using surfaces with two independently tunable length scales, we isolate this near-wall coupling from a second collective process, coalescence between departing bubble clusters, which transitions through isolated, promotive, and excessive regimes as the departure diameter grows with heat flux. The dominant length scale thus shifts with operating conditions, from boundary-layer coupling near activation to departure-scale coupling once nucleation is established. These results establish a scale-dependent framework for collective nucleation and departure, broadly related to phase change processes on structured surfaces.

## Introduction

Nucleation is the emergence of a new phase from a metastable medium, which underpins many phase-change phenomena [1–3], yet how it establishes, persists, and varies on different surfaces often remains poorly understood [4,5]. In boiling, this gap is particularly consequential: heat transfer efficiency relies on a delicate balance between bubble generation near the surface and removal into the bulk, making reliable nucleation a prerequisite for robust performance across applications from power generation [6] to electronics cooling [7] and chemical processing [8]. Increasingly sophisticated surface engineering strategies have enabled marked boiling performance enhancement by tuning surface geometries [9,10], wettabilities [11], and liquid supply pathways [12,13]. However, even carefully micro-engineered, nominally identical surfaces often exhibit large variations in activation superheat, nucleation persistence, and hysteresis under changing thermal loads, leading to significant variability in heat transfer across the nucleate boiling regime [14–16]. Such variability points to a fundamental uncertainty in how nucleation emerges and persists at the surface, which can directly constrain reliability, control strategies, and design margins in scenarios where boiling must operate robustly rather than just optimally. Nevertheless, most boiling studies assess performance primarily through the critical heat flux and heat transfer coefficients measured after nucleation is fully established [9], leaving the role of nucleation stability and variability in boiling heat transfer unresolved.

Classically, nucleation and its variability are interpreted through the activation behavior of individual surface features embedded within the thermal boundary layer [17,18]. Differences in superheat for site activation are commonly attributed to different surface microstructures [19], vapor entrapment [15,20], and thermal history [14]. Within this view, the near-wall thermal and flow field is treated as a prescribed background condition set by the applied heat flux and bulk liquid environment, while nucleation sites are assumed to respond independently to this field. This single-site framework has been successful in rationalizing trends in activation thresholds. However, it does not capture the possibility that neighboring sites collectively reshape the thermal and flow conditions that govern each site's activation and persistence. Whether such geometric coupling plays a significant role in nucleation stability remains an open question.

In most previous studies, interactions between neighboring sites have been investigated in the context of bubble growth and departure, instead of nucleation onset and persistence. Bubble departure frequency and size are typically measured as a function of the site spacing normalized by the buoyancy-driven departure diameter $D_{\mathrm{buo}}$ [15,21–27]. Several interaction mechanisms have been proposed, including thermal interactions mediated by overlapping near-wall temperature fields [22,26,27], hydrodynamic interactions driven by liquid motion during bubble growth and departure [22,26,27], and bubble coalescence that alters interfacial energy and departure dynamics [22,26–30]. The few studies that specifically examined inter-site effects on nucleation activation attributed them to substrate thermal diffusion [31] and vapor-mediated seeding of adjacent sites [21,31,32], but without recognizing a near-wall hydrodynamic length scale distinct from the bubble departure scale. Nor did they systematically characterize nucleation variability and persistence and their impact on heat transfer. As a result, near-wall processes governing nucleation stability and larger-scale processes governing vapor removal have been studied largely in isolation, without a unifying framework linking them through the geometric length scales at which they operate.

Two fundamental knowledge gaps limit progress toward such a framework: first, how geometric coupling between neighboring sites influences nucleation activation, deactivation, and stability; and second, how interactions at different length scales contribute distinctly to overall heat transfer across the nucleate boiling regime. Addressing both requires recognizing that boiling operates across at least two physically distinct length scales (**Fig. 1a**). Near the wall, nucleation onset and persistence are governed by the thermal/hydrodynamic boundary layer thickness (thermal boundary layer thickness $\delta_{\mathrm{th}} \approx 206$ μm and hydrodynamic boundary layer thickness $\delta_{\mathrm{H}} \approx 248$ μm in our experiments, **Fig. S1** and **Supplementary Note S1**); at larger scales, bulk vapor removal is governed by the bubble departure diameter $D_{\mathrm{b}}$. Since bubble coalescence during growth can alter departure dynamics [28,29,33], $D_{\mathrm{b}}$ does not necessarily equal the buoyancy-driven diameter $D_{\mathrm{buo}}$ (≈ 2.5 mm in our experiments) of an isolated bubble. Further, in many prior experiments, interacting bubbles either occupied only a small fraction of the heated surface, yielding limited thermal signal-to-noise ratio or spanned the entire substrate in an uncontrolled manner, obscuring any clear separation between local near-wall interactions and larger-scale hydrodynamic effects.

In this study, we design a surface that explicitly decouples $\delta_{\mathrm{H}}$-scale and $D_{\mathrm{b}}$-scale interactions (**Fig. 1b**). Our surface consists of eight cylindrical cavities organized into four discrete pairs, thereby defining two independent geometric parameters: the intra-pair spacing $L_{\mathrm{in}}$ between immediate neighbors, varied at the scale of $\delta_{\mathrm{H}}$ and inter-pair separation, characterized by the diagonal distance $L_{\mathrm{D}}$ between cavity pairs, varied at the scale of $D_{\mathrm{b}}$. This dual-scale architecture enables controlled investigation and high-speed visualization of near-wall nucleation interactions together with larger-scale vapor removal dynamics. We show that when $L_{\mathrm{in}}$ approaches $\delta_{\mathrm{H}}$, nucleation becomes both easier to initiate and more resistant to suppression. This behavior is consistent with a hydrodynamic shielding mechanism: neighboring bubbles generate a low-velocity, locally overheated region between sites that reduces convective heat loss and stabilizes vapor embryos. In contrast, once all sites are activated, the overall heat transfer becomes largely insensitive to $L_{\mathrm{in}}$, and hence to near-wall intra-pair interactions; instead, it is governed by macroscopic entrainment and inter-pair coalescence patterns determined by $L_{\mathrm{D}}/D_{\mathrm{b}}$ [22,25]. Together, these results demonstrate a clear separation between near-wall hydrodynamic interactions that regulate nucleation stability and macroscale mechanisms that govern vapor removal and heat transfer, with $L_{\mathrm{in}}/\delta_{\mathrm{H}}$ and $L_{\mathrm{D}}/D_{\mathrm{b}}$ emerging as the natural design parameters for boiling surfaces.

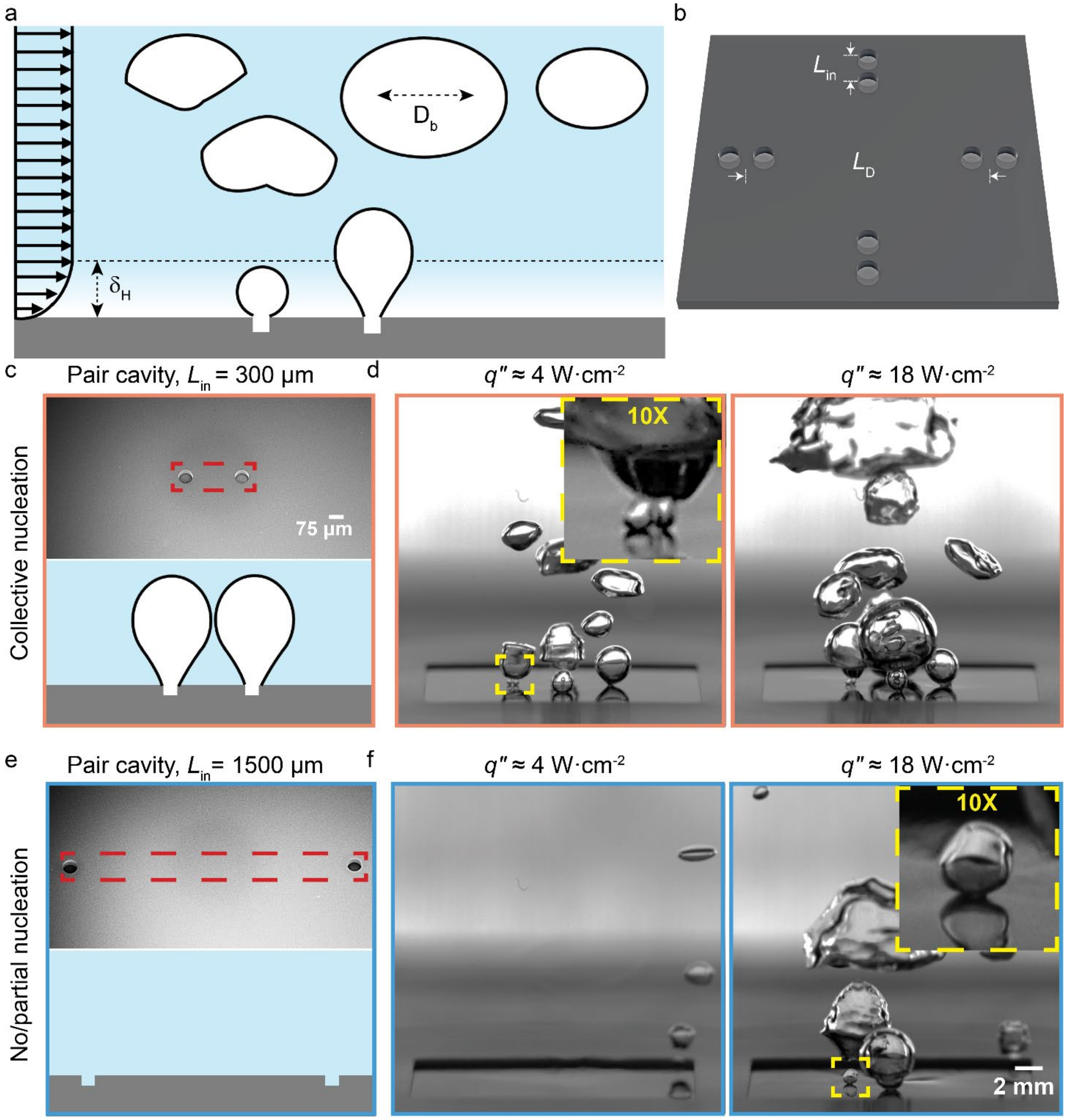

**Fig. 1 Dual-scale nucleation-site layout and contrasting boiling behavior between closely and widely spaced cavities.** (a) Schematic illustrating the two scales governing nucleation site interactions: the hydrodynamic boundary layer thickness $\delta_H$ near the heated wall and the bubble departure diameter $D_b$. Arrows indicate the natural convection velocity profile. (b) Schematic of the micro-engineered surface consisting of eight identical cylindrical cavities arranged into four discrete pairs. The geometry defines two characteristic length scales: the intra-pair spacing $L_{in}$ between neighboring cavities and the inter-pair characteristic length $L_D$ given by the diagonal distance between cavity pairs. Cavity diameter and depth are fixed at 75 μm and 50 μm, respectively. (c) Top-view SEM image of a closely spaced cavity pair ($L_{in}$ = 300 μm) and schematic showing collective nucleation. (d) Side-view high-speed images for $L_{in}$ = 300 μm, $L_D$ = 7000 μm at $q'' \approx 4$ and 18 W·cm$^{-2}$ during a decreasing heat flux sweep. At low heat flux, all cavities remain active (yellow inset: 10× magnification of the nucleation region). (e) Top-view SEM image of a widely spaced cavity pair ($L_{in}$ = 1500 μm) and schematic showing no or partial nucleation. (f) Side-view high-speed images for $L_{in}$ = 1500 μm, $L_D$ = 7000 μm at $q'' \approx 4$ and 18 W·cm$^{-2}$ during a decreasing heat flux sweep. At low heat flux, no nucleation activity is observed at any cavity. At moderate heat flux, bubbles emerge only from a subset of predefined nucleation sites (yellow inset: 10× magnification).

## Results

We first examine how intra-pair spacing $L_{in}$ governs nucleation behavior while limiting inter-pair interactions by fixing $L_D$ at 7000 μm. Two representative configurations are compared: closely spaced cavities with $L_{in}$ = 300 μm, comparable to $\delta_H$, and widely spaced cavities with $L_{in}$ = 1500 μm, about six times larger than $\delta_H$. All other cavity parameters are identical: 75 μm diameter, 50 μm depth, thermally grown $SiO_2$ on the top surface, and Si cavity walls etched by deep reactive-ion etching. Full microfabrication procedures are provided in the Methods (also see **Figs. S2** and **S3** and **Supplementary Notes S2** and **S3**).

**Figure 1c-1f (Supplementary Videos 1 and 2)** contrast nucleation behavior of the two configurations. Panels c and e show SEM images alongside schematics of the observed nucleation behavior; panels d and f present the corresponding side-view high-speed snapshots at a low heat flux ($q'' \approx 4$ W·cm⁻²) and a moderate heat flux ($q'' \approx 18$ W·cm⁻²). For the closely spaced configuration ($L_{in}$ = 300 μm), the schematic in **Fig. 1c** illustrates "*collective nucleation*", where both neighboring cavities are active and interact within a shared near-wall region. The high-speed snapshots in **Fig. 1d** experimentally demonstrate this picture: even at the low heat flux of $q'' \approx 4$ W·cm⁻², synchronized bubble growth is already visible from neighboring sites (magnified inset).

The widely spaced configuration ($L_{in}$ = 1500 μm) exhibits markedly different behavior at low-to-moderate heat fluxes, which we term "*no/partial nucleation*" (**Fig. 1e**). At $q'' \approx 4$ W·cm⁻², no bubble activity is observed at any of the engineered cavities (occasional bubbles near the chip corners originate from peripheral regions outside the patterned cavity array and heating region, which do not affect the reported trends). Even at $q'' \approx 18$ W·cm⁻², bubble nucleation is confined to a subset of cavities; the inset in **Fig. 1f** shows a single bubble growing and departing independently from one cavity, while its neighbor located 1500 μm away remains inactive. This contrast, produced by $L_{in}$ alone on an otherwise identical surface, points to $\delta_H$ as the governing length scale for near-wall site interactions.

The snapshots in **Fig. 1** were obtained in our custom-built saturated pool-boiling apparatus (**Fig. 2**), which provides continuous side-view high-speed visualization of the artificial nucleation sites through a transparent glass chamber filled with deionized water. The micro-engineered test substrate (2 cm × 2 cm) is mounted within the chamber using precision-machined PEEK holders and mechanical clamping, eliminating adhesives that can introduce surface contamination and outgassing during repeated experimental cycles. Beyond visualization, this setup also enables accurate boiling heat transfer characterization: on the backside of the substrate, we microfabricate a Pt serpentine thin-film heater (1 cm × 1 cm; **Fig. 2** inset) that simultaneously serves as a Joule heater and a resistance temperature detector (RTD). Together, the on-chip heater/RTD and high-speed visualization allow us to quantify nucleation activation, persistence, and overall heat-transfer performance across different configurations.

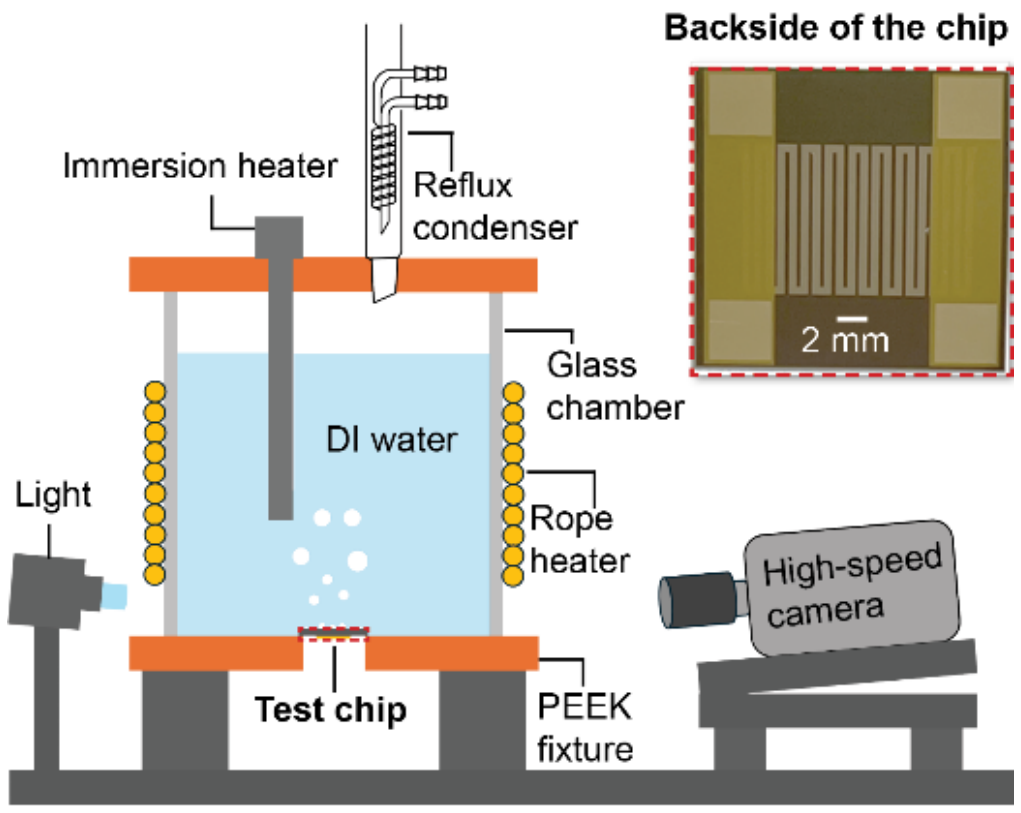

**Fig. 2 Experimental setup for pool-boiling measurements and visualization**. The micro-engineered test surface, integrated with an on-chip Pt thin-film heater that also serves as an RTD (inset), is mounted at the base of a transparent

glass chamber filled with deionized water. The bulk liquid is maintained at the saturation temperature (100 °C) at ambient pressure using an immersion heater and a surrounding rope heater, while a reflux condenser minimizes net vapor loss. Bubble nucleation and dynamics are visualized through the side wall of the chamber using a high-speed camera with external backlighting.

During each experiment, we record the applied voltage and measure the current across the Pt heater, yielding the input power and the instantaneous electrical resistance (**Fig. S4** and **Supplementary Note S4**). This resistance is subsequently converted to the mean heater temperature via chip-specific calibration curves (**Fig. S5** and **Supplementary Note S5**). As the 1 cm × 1 cm heating area is embedded in the 2 cm × 2 cm chip (**Fig. 2** inset), we account for lateral heat spreading and peripheral convection losses using a steady-state heat-transfer simulation with COMSOL Multiphysics. Applying this correction to each data point returns the boiling heat flux $q''$ and sample temperature $T$ averaged over the 1 cm² heater area (**Fig. S6** and **Supplementary Note S6**). To construct the boiling curves, at each heat-flux step, we hold the system for at least 10 min before data acquisition to ensure steady-state conditions. Full experimental procedures and modeling details are provided in the Methods.

## Nucleation-site occupancy and near-wall hydrodynamic interactions

To directly probe how intra-pair spacing $L_{in}$ governs collective nucleation, we track $N$, the number of active nucleation sites among the eight engineered cavities, as a function of heat flux. After thoroughly degassing the system (see Methods), we perform activation sweeps (increasing $q''$ in steps) until all sites are active, followed by deactivation sweeps (decreasing $q''$ in steps) back to the single-phase regime. Mapping $N$ as a function of $q''$ for both activation and deactivation sweeps resolves the heat flux range in which nucleation initiates, how many sites become active, and whether activity is retained when $q''$ is reduced. **Figure 3a** and **3b** present these maps for $L_{in}$ = 300 and 1500 μm: panel a shows the activation sweep and panel b the deactivation sweep. Each grid cell represents a discrete heat-flux bin; only recorded snapshots in which the activation state of every cavity could be unambiguously resolved are included. Color intensity encodes how often that ($N$, $q''$) state was observed across repeated sweeps and different samples, with darker shading indicating a more frequently observed state.

During activation sweeps (**Fig. 3a**), closely spaced cavity pairs ($L_{in}$ = 300 μm, red) reach higher site occupancy at significantly lower $q''$ compared to widely spaced pairs ($L_{in}$ = 1500 μm, blue). As $q''$ increases, the 300 μm configuration transitions rapidly from partial to full occupancy ($N$ = 8) at low-to-moderate heat flux, whereas the 1500 μm configuration activates progressively and only reaches full activation at substantially higher $q''$. During deactivation sweeps (**Fig. 3b**), the contrast sharpens further: the 300 μm configuration retains all eight active cavities down to low $q''$; meanwhile, the 1500 μm configuration begins losing active sites at much higher $q''$ and at each heat-flux bin, the observed $N$ is spread across a wide range of values rather than clustered at a single level. Because all cavities share identical geometry, surface chemistry, and global thermal boundary conditions (**Fig. S1**), these systematic differences isolate the intra-pair cavity spacing, and specifically the near-wall interactions it governs, as the controlling variable.

A plausible physical origin of these near-wall interactions is illustrated in **Fig. 3c** through representative two-dimensional flow-field simulations of liquid motion around neighboring surface bubbles at close spacing (top) and wide spacing (bottom). When $L_{in} \sim \delta_H$, the flow speed between adjacent bubbles is substantially suppressed. This hydrodynamic shielding reduces convective heat removal from the inter-cavity region, stabilizing vapor embryos and collectively lowering the effective activation threshold. On the other hand, for widely spaced bubbles ($L_{in} \approx 6\delta_H$), each bubble's near-wall flow is largely unaffected by its neighbor, and each site experiences a more independent thermal and fluidic environment. Consequently, collaborative stabilization is absent, and sites activate and deactivate individually. Therefore in **Fig. 3a** and **3b**, we observe cooperative activation and robust retention at $L_{in}$ = 300 μm, versus delayed activation and earlier, more stochastic deactivation at $L_{in}$ = 1500 μm. These simulations in **Fig. 3c**, while two-dimensional, capture the essential near-wall velocity contrast, consistent with the occupancy trends in panels a and b (see **Supplementary Note S7** for full simulations details).

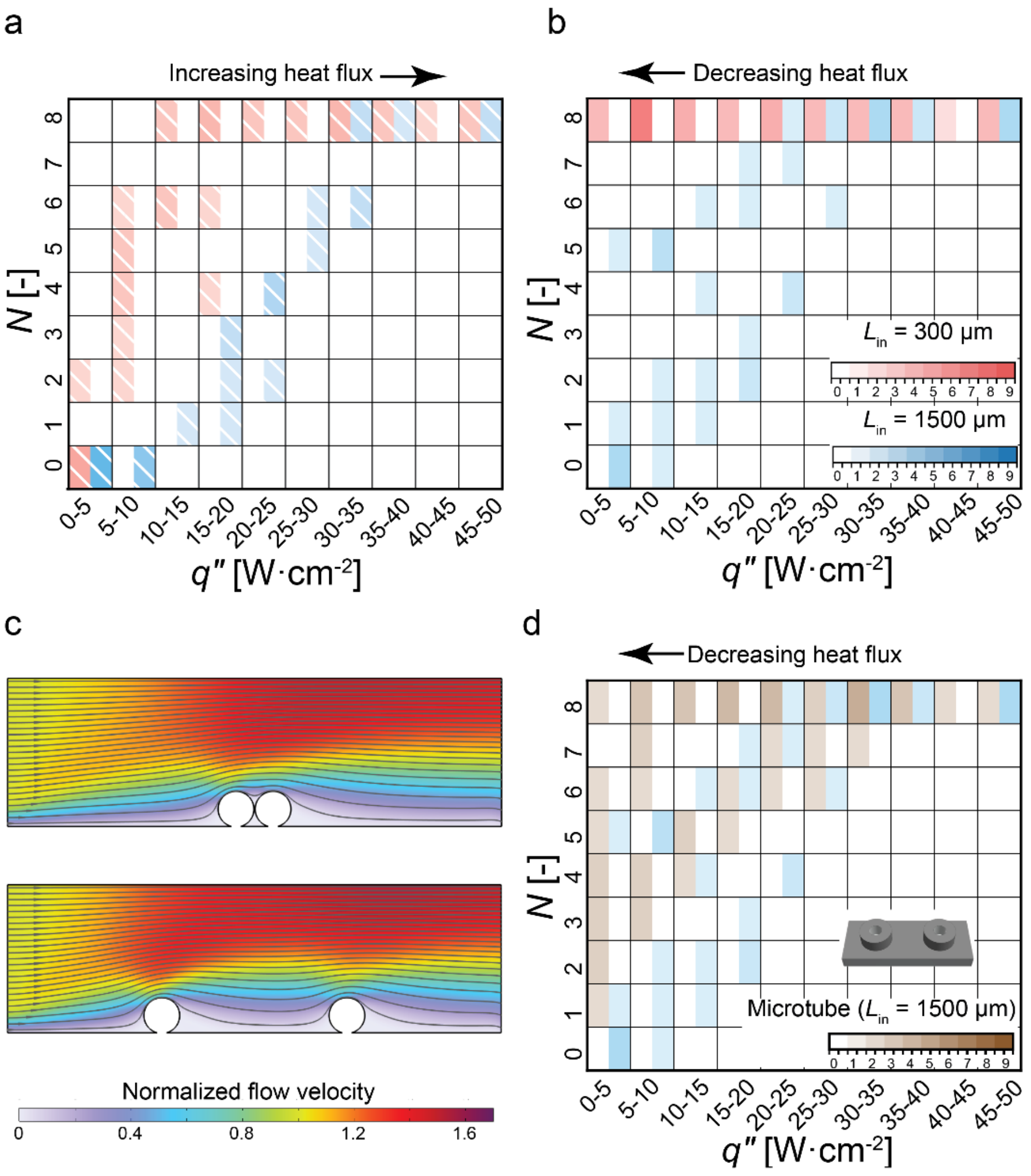

**Fig. 3 Nucleation-site occupancy and hydrodynamic shielding mechanism.** (a) Activation sweeps and (b) deactivation sweeps showing the number of active nucleation sites, $N$, out of eight engineered cavities, as a function of applied heat flux $q''$. In (a), hatched bins denote increasing $q''$; in (b), solid bins denote decreasing $q''$. Each grid cell represents a discrete heat-flux bin, with color intensity encoding how often that ($N$, $q''$) state was observed across repeated sweeps and samples; darker shading indicates higher observation frequency (only frames in which the activation state of every cavity could be unambiguously resolved are included). Closely spaced cavity pairs ($L_{in}$ = 300 μm, red) activate and reach full site occupancy at lower heat flux than widely spaced pairs ($L_{in}$ = 1500 μm, blue) while retaining full occupancy down to significantly lower heat fluxes during deactivation. (c) Two-dimensional numerical simulations of near-wall liquid velocity around adjacent bubbles. When $L_{in} \sim \delta_H$ (top), the flow between the bubbles is substantially slowed, producing a shielded region at the wall; when $L_{in}$ is well above $\delta_H$ (bottom), each bubble's near-wall environment is largely unaffected by its neighbor. (d) Deactivation map for microtubes (500 μm outer diameter, 75 μm inner diameter, 50 μm height; brown) at $L_{in}$ = 1500 μm, compared with the planar cavity data at the same $L_{in}$ (blue). The elevated tube walls physically obstruct near-wall flow, producing effective shielding even at wide intra-pair spacing: as a result, microtubes retain active sites down to lower $q''$ than planar cavities at identical $L_{in}$.

To test whether flow obstruction is indeed the controlling factor, we designed a surface on which each nucleation site is surrounded by a raised wall that physically blocks near-wall liquid flow, independent of any neighboring bubble. Specifically, we replaced each planar cavity with a microtube, a 50 μm tall pillar housing a nucleation site (500 μm outer diameter, 75 μm inner diameter). The elevated tube walls obstruct the near-wall flow around each site directly, producing a shielding effect even when the intra-pair spacing is large. **Figure 3d** compares nucleation persistence during decreasing heat flux sweeps for microtube pairs ($L_{in}$ = 1500 μm, brown) against planar cavity pairs with the same $L_{in}$ (blue). Despite identical intra-pair spacing, the microtube configuration retains active sites down to much lower $q''$, consistent with the enhanced geometric obstruction of near-wall liquid flow. We note that microtubes are sometimes associated with improved gas entrapment, but this effect is minimized here as we analyze deactivation sweeps after prolonged degassing and sustained boiling, starting from a full-activation state. Moreover, comparing **Fig. 3b** to **Fig. 3d**, the flat paired-cavity surface with $L_{in}$ = 300 μm (red) shows greater nucleation persistence than microtubes with $L_{in}$ = 1500 μm, which indicates that differences in gas or vapor entrapment alone are insufficient to account for the observed trends, reinforcing the role of near-wall hydrodynamic interactions.

## Nucleation-dominated heat transfer performance and variability

Having established how intra-pair spacing governs site activation and retention (**Fig. 3**), we next quantify its thermal consequences. **Figure 4** records boiling curves ($q''$ vs. $T$) through controlled heat-flux ramps on the same $L_{in}$ = 300 and 1500 μm paired geometries (with $L_D$ fixed at 7000 μm). Panel a reports activation sweeps and panel b deactivation sweeps; each aggregates multiple runs on the same sample and across nominally identical samples, so the spread at a given $q''$ reflects genuine run-to-run and sample-to-sample variability independent of fixed measurement error from instrumentation (±0.3 °C in $T$; ±0.1 W·cm⁻² in $q''$, **Supplementary Note S8**).

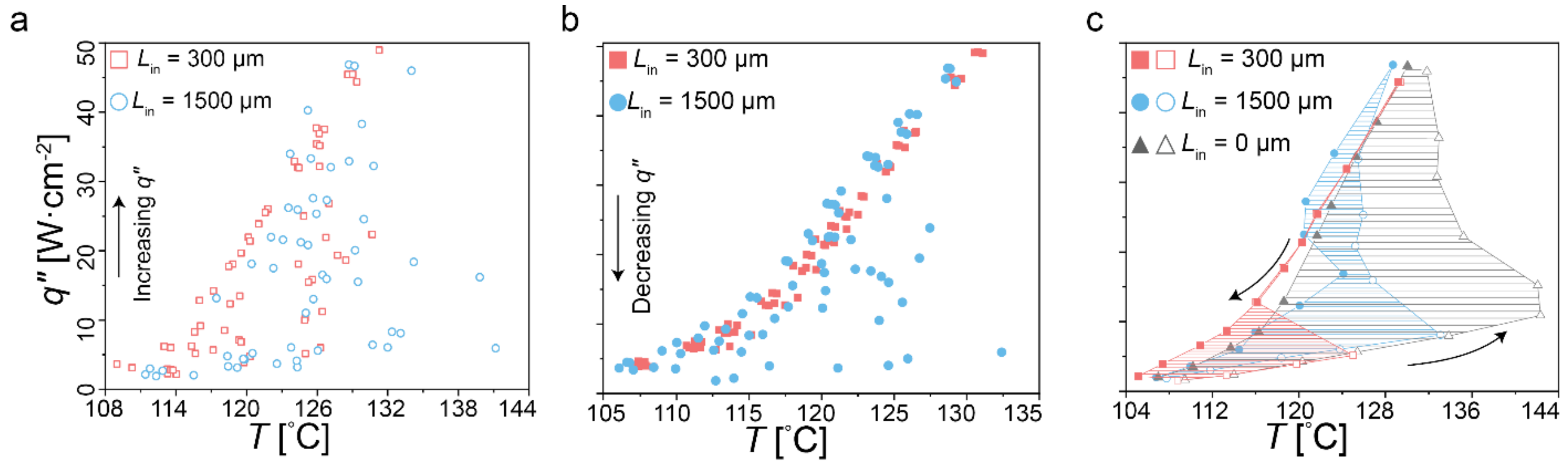

**Fig. 4 Nucleation-dominated boiling heat transfer.** (a) Heat flux $q''$ vs temperature $T$ during increasing heat-flux sweeps (activation) for paired cavities with $L_{in}$ = 300 μm (red squares) and 1500 μm (blue circles). Both geometries show scatter at a given $q''$, markedly greater for widely spaced pairs. (b) $q''$ vs $T$ during decreasing heat-flux sweeps (deactivation) for the same geometries. Closely spaced cavity pairs exhibit sharply defined, reproducible deactivation behavior; widely spaced pairs show pronounced run-to-run and sample-to-sample variability. (c) Representative activation (open symbols) and deactivation (solid symbols) trajectories for three configurations ($L_{in}$ = 300 μm, red squares; 1500 μm, blue circles; 0 μm, unpaired cavities, gray triangles), each from a single sample. Shaded regions indicate the hysteresis envelope between two sweep directions. All cases converge to similar behavior at high heat flux ($q''$ ≈ 45 W·cm$^{-2}$). Inter-pair separation $L_D$ is fixed at 7000 μm for all data shown. At each heat-flux step, the system is held for at least 10 min prior to data acquisition to ensure steady-state conditions. The fixed measurement uncertainties of $T$ and $q''$ for each data point are 0.3 °C and 0.1 W·cm$^{-2}$, respectively.

During activation sweeps (**Fig. 4a**, open symbols), both $L_{in}$ = 300 and 1500 μm samples exhibit noticeable scatter in $T$ at a given $q''$, consistent with the stochastic nature of nucleation onset (also observed in **Fig. 3a**). Widely spaced pairs (blue circles) scatter more significantly than closely spaced pairs (red squares), which is a direct thermal manifestation of the site-by-site activation documented in **Fig. 3a**: when cavities activate independently, the boiling onset is less predictable and the activation temperature is on average higher. This contrast sharpens during deactivation sweeps (**Fig. 4b**, closed symbols). The $L_{in}$ = 300 μm configuration yields tightly clustered, highly reproducible trajectories, reflecting the stable, full-occupancy deactivation behavior in **Fig. 3b**. The $L_{in}$ = 1500 μm configuration, on the other hand, displays substantial run-to-run and sample-to-sample variability, due to its variable and incomplete site retention during decreasing heat flux sweeps. Note that these $L_{in}$-dependent trends are not specific to a particular $L_D$ (see **Fig. S7** for $L_D$ = 5000 μm).

**Figure 4c** overlays representative activation and deactivation trajectories for the two paired geometries and an unpaired four-single-cavity reference ($L_{in}$ = 0 μm, gray triangles) to visualize hysteresis through the shaded region between the two sweeps. The closely spaced pairs ($L_{in}$ = 300 μm) exhibit narrower hysteresis compared to the widely spaced pairs ($L_{in}$ = 1500 μm), as $\delta_H$-scale site interactions both lower the activation superheat and sustain nucleation through deactivation. In addition, the unpaired configuration ($L_{in}$ = 0 μm) shows the largest hysteresis of the three, further confirming that neighbor proximity is necessary for collective nucleation behavior. Interestingly, despite these pronounced spacing-dependent differences at low and moderate heat fluxes, all three configurations converge toward similar thermal behavior at high heat fluxes ($q''$ ≈ 45 W·cm$^{-2}$), indicating that once all sites are fully active, heat transfer is governed by mechanisms beyond nucleation stability. We examine this post-activation regime in the next section.

## Vapor removal and heat transfer after full site activation

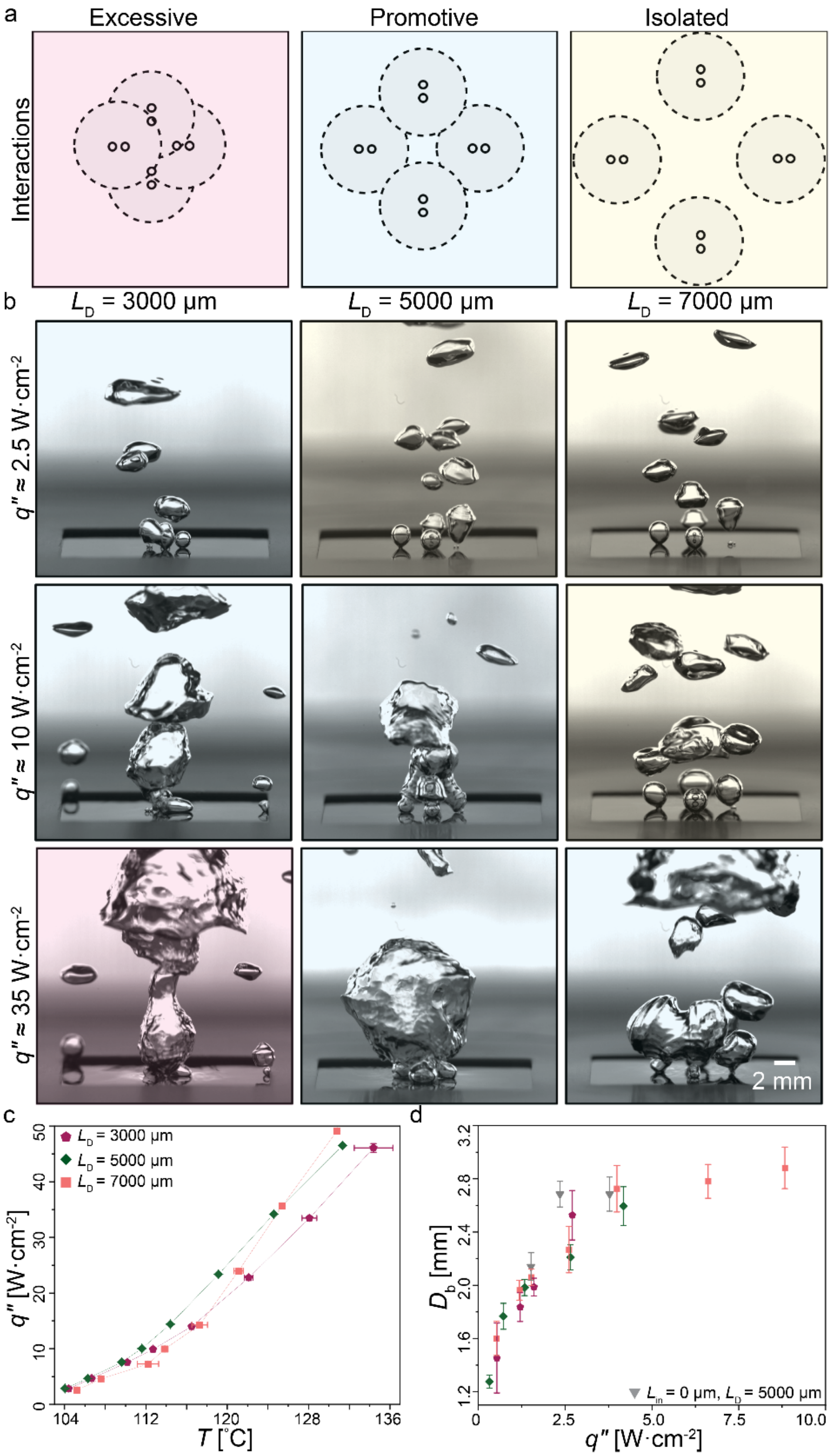

**Fig. 5 Vapor removal governed by inter-pair spacing after full activation.** (a) Schematics of three inter-pair bubble interaction regimes defined by $L_D/D_b$: excessive coalescence (small $L_D/D_b$, pink), where merged vapor plumes impede liquid replenishment; promotive interaction (intermediate $L_D/D_b$, yellow), where coalescence or entrainment between adjacent departing clusters enhances vapor removal; and isolated departure (large $L_D/D_b$, blue), where vapor clusters

from each pair behave independently. (b) Representative side-view images of bubble dynamics at $q'' \approx 2.5$, 10, and 35 $W \cdot cm^{-2}$ for $L_D$ = 3000, 5000, and 7000 μm with $L_{in}$ fixed at 300 μm during a deactivation sweep. Background colors indicate the prevailing interaction regime at each combination of $L_D$ and $q''$ (color scheme as in panel a). **(c)** $q''$ vs $T$ during deactivation sweeps at full site occupancy ($N$ = 8) for $L_D$ = 3000 μm (purple pentagons), 5000 μm (green diamonds), and 7000 μm (red squares). (d) $D_b$ vs $q''$ prior to inter-pair bubble coalescence for $L_D$ = 3000, 5000, and 7000 μm; gray triangles show an unpaired four-single-cavity reference ($L_{in}$ = 0 μm, $L_D$ = 5000 μm). Error bars in panels c and d reflect combined statistical and measurement uncertainties.

Having shown how $\delta_H$ governs nucleation onset and stability, we now investigate how inter-pair separation $L_D$ controls vapor removal once all sites are fully active. With $L_{in}$ fixed at 300 μm to maintain full site occupancy ($N$ = 8) throughout deactivation sweeps (**Fig. 3b** and **Fig. 4b**), we compare three inter-pair separations: $L_D$ = 3000, 5000, and 7000 μm. Depending on $L_D/D_b$, neighboring bubble clusters may interact in three distinct ways (**Fig. 5a**): at small $L_D/D_b$, merged vapor masses obstruct liquid replenishment (excessive, pink); at intermediate $L_D/D_b$, mutual entrainment or coalescence between departing clusters enhances vapor removal (promotive, yellow); at large $L_D/D_b$, pairs depart independently (isolated, blue). Since $D_b$ grows with $q''$, a fixed geometry can traverse multiple regimes as heat flux rises.

At low heat flux ($q'' \approx 2.5$ $W \cdot cm^{-2}$; **Fig. 5b**, top row, **Supplementary Video 3**), inter-cluster coalescence has already begun for $L_D$ = 3000 μm, placing it in the promotive regime while bubble clusters for $L_D$ = 5000 and 7000 μm remain isolated. The heat transfer consequences are modest at this heat flux (**Fig. 5c**): $L_D$ = 3000 and 5000 μm (purple pentagons and green diamonds) show comparable thermal advantage over $L_D$ = 7000 μm (red squares), attributable to promotive coalescence at $L_D$ = 3000 μm and wake-enhanced departure at $L_D$ = 5000 μm [22], respectively.

As $q''$ increases, $D_b$ grows monotonically across all configurations (**Fig. 5d**), so the effective spacing between departing clusters ($L_D/D_b$) shrinks. Here, $D_b$ is extracted using BubbleID [34], an automated image-analysis algorithm that identifies and tracks individual bubble contours frame-by-frame (**Supplementary Note S9**). At $q'' \approx 10$ $W \cdot cm^{-2}$ (**Fig. 5b**, middle row, **Supplementary Video 4**), promotive interactions have intensified at $L_D$ = 3000 μm, and $L_D$ = 5000 μm has entered the promotive regime as $D_b$ has grown. $L_D$ = 5000 μm achieves the lowest surface temperature in **Fig. 5c** at this heat flux. At $L_D$ = 7000 μm (blue background), however, $D_b$ remains small relative to $L_D$ and pairs continue to depart largely in the isolated regime.

At $q'' \approx 35$ $W \cdot cm^{-2}$ (**Fig. 5b**, bottom row, **Supplementary Video 5**), $L_D$ = 3000 μm has transitioned into the excessive regime: a coalesced vapor mass now spans multiple pairs and obstructs liquid replenishment, driving its boiling curve to the highest surface temperature in **Fig. 5c**. At $L_D$ = 7000 μm, $D_b$ has grown large enough to reach promotive coalescence between neighboring clusters, and its boiling curve converges toward that of $L_D$ = 5000 μm, which remains in the promotive regime. Overall, the regime a geometry occupies can shift with $q''$; $L_D/D_b$, not $L_D$ alone, is therefore the relevant parameter, and the optimal inter-pair spacing depends on the intended operating heat flux.

## Unifying nucleation and vapor-removal controls across heat flux

On a real boiling surface, nucleation sites may be distributed across separations that span both $\delta_H$ and $D_b$ scales. **Figure 6** isolates the action of each length scale through two designed comparisons. Holding $L_{in}$ fixed at 300 μm, we compare $L_D$ = 3000 μm (purple pentagons) and $L_D$ = 7000 μm (red squares) to probe inter-pair effects at a common intra-pair spacing; holding $L_D$ fixed at 7000 μm, we compare paired cavities ($L_{in}$ = 300 μm) against an unpaired four-single-cavity reference ($L_{in}$ = 0 μm, gray triangles) to probe intra-pair effects at a common inter-pair separation. **Figure 6a** records boiling curves during deactivation sweeps; error bars reflect the combined measurement and statistical uncertainty, dominated by run-to-run and sample-to-sample variability.

Which spacing controls the boiling curves depends on heat flux: at low $q''$, the curves are separated by $L_{in}$ (paired vs. unpaired) while the two paired curves remain close despite the more than two-fold difference in $L_D$; at high $q''$, the two paired curves diverge by $L_D$, whereas the unpaired curve converges toward its paired counterpart at the same $L_D$. The boiling curves themselves thus illustrate a handoff: $L_{in}$ governs the low-$q''$ response, $L_D$ governs the high-$q''$ response.

At low heat flux ($q'' \approx 2.5$ W·cm$^{-2}$), the two paired configurations ($L_{in}$ = 300 μm) both exhibit collective nucleation at all sites (**Fig. 6b**, top row, purple and red boxes); the slight heat transfer difference between them at this heat flux is due to wake-assisted bubble departure for $L_D$ = 3000 μm (see **Fig. 5b**). The unpaired configuration ($L_{in}$ = 0 μm), lacking intra-pair neighboring cavities to provide hydrodynamic shielding, produces larger temperature uncertainties and on average requires higher superheat to sustain nucleate boiling.

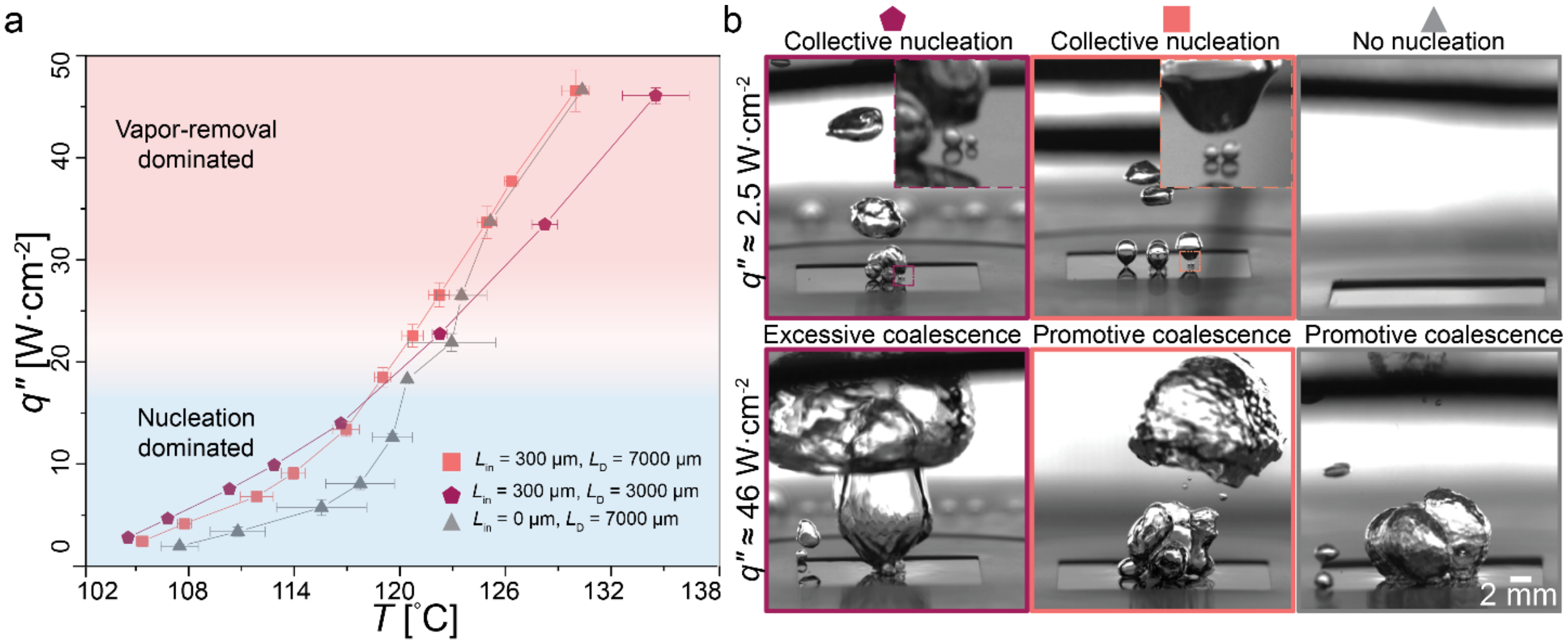

**Fig. 6 Heat-flux-dependent handoff between $L_{in}$- and $L_D$-controlled regimes.** (a) $q''$ vs $T$ and (b) corresponding high-speed snapshots for two paired configurations ($L_{in}$ = 300 μm with $L_D$ = 3000 and 7000 μm) and an unpaired four-single-cavity reference ($L_{in}$ = 0 μm, $L_D$ = 7000 μm), obtained during deactivation sweeps. At low heat flux, the unpaired four-single-cavity reference shows no nucleation and the two paired configurations exhibit collective nucleation at all sites, confirming that $L_{in}$ controls nucleation stability. At high heat flux, the two $L_D$ = 7000 μm curves converge, while the $L_D$ = 3000 μm surface shifts to higher $T$ due to excessive inter-pair coalescence and impeded liquid replenishment, confirming that $L_D$ governs vapor removal efficiency.

As $q''$ increases and all configurations reach full site occupancy, a large inter-pair separation leads to promotive vapor removal regardless of nucleation history. At high heat flux ($q'' \approx 46$ W·cm$^{-2}$), the unpaired and paired $L_D$ = 7000 μm surfaces both occupy the promotive regime (**Fig. 6b**, bottom row, red and gray boxes), confirming that $L_{in}$ no longer controls the response. In contrast, the paired $L_D$ = 3000 μm configuration enters the excessive coalescence regime with sprawling vapor structures that depart intermittently (**Fig. 6b**, bottom row, purple box), shifting its boiling curve to the highest surface temperature and largest variability. Therefore, the controlling spacing for boiling performance depends on the operating heat flux, $L_{in}$ at low $q''$ and $L_D$ at high $q''$.

## Discussion

The heat-flux-dependent handoff reflects two distinct hydrodynamic length scales that act at different stages of boiling. Intra-pair spacing controls near-wall hydrodynamic shielding: when $L_{in}$ is comparable to the hydrodynamic boundary-layer thickness $\delta_H$, collective modification of the near-wall flow by neighboring bubbles stabilizes vapor embryos, lowers the effective activation threshold, sustains nucleation during deactivation, and suppresses boiling hysteresis. The microtube data (**Fig. 3d**) corroborates this physical picture: elevated tube walls further restrict near-wall liquid flow and therefore yield more persistent nucleation. Direct experimental confirmation of the modified near-wall flow during active nucleation remains technically challenging and a direction for future work that would enable quantitative characterization of regime boundaries.

Inter-pair separation, by contrast, governs vapor removal through its relation to the departure diameter $D_b$: depending on $L_D/D_b$, adjacent bubble plumes may remain isolated, coalesce promotively, or coalesce excessively and obstruct liquid return. These three regimes have often been treated separately in prior literature: excessive coalescence primarily in the context of dryout and critical heat flux, and promotive coalescence and independent bubble growth in studies of site interactions. Because $D_b$ grows with $q''$, one surface can traverse multiple $L_D/D_b$ regimes as operating conditions shift, unifying these previously separate observations under a single framework.

Overall, the two controlling length scales, $\delta_H$ for near-wall shielding and $D_b$ for interactions between departing bubbles, are set by fluid properties and working conditions, making the regime behavior broadly applicable beyond the centimeter-scale water/silicon system studied here. Our work reveals that the dominant interaction length scale in boiling evolves with thermal loads, a perspective that is essential for the design of predictable, high-performance phase change surfaces across operating conditions.

## References

1. Dhir, V. K. Boiling Heat Transfer. *Annual Review of Fluid Mechanics* vol. 30 365–401 (1998).
2. Mandsberg, N. K. Spatial control of condensation: the past, the present, and the future. *Advanced Materials Interfaces* **8**, 2100815 (2021).
3. Nath, S., Ahmadi, S. F. & Boreyko, J. B. A review of condensation frosting. *Nanoscale and Microscale Thermophysical Engineering* **21**, 81–101 (2017).
4. Lohse, D. Bubble puzzles: From fundamentals to applications. *Phys. Rev. Fluids* **3**, 110504 (2018).
5. Dhir, V. K. Mechanistic Prediction of Nucleate Boiling Heat Transfer–Achievable or a Hopeless Task? *Journal of Heat Transfer* **128**, 1–12 (2005).
6. Theriault, K. Boiling water reactors. in *Nuclear Engineering Handbook* 85–140 (CRC Press, 2016).
7. Fan, S. & Duan, F. A review of two-phase submerged boiling in thermal management of electronic cooling. *International Journal of Heat and Mass Transfer* **150**, 119324 (2020).
8. Chemat, F. & Esveld, E. Microwave Super-Heated Boiling of Organic Liquids: Origin, Effect and Application. *Chem. Eng. Technol.* **24**, 735–744 (2001).
9. Cho, H. J., Preston, D. J., Zhu, Y. & Wang, E. N. Nanoengineered materials for liquid–vapour phase-change heat transfer. *Nat Rev Mater* **2**, 16092 (2016).
10. Jaikumar, A. & Kandlikar, S. G. Ultra-high pool boiling performance and effect of channel width with selectively coated open microchannels. *International Journal of Heat and Mass Transfer* **95**, 795–805 (2016).
11. Betz, A. R., Jenkins, J., Kim, C.-J. "Cj" & Attinger, D. Boiling heat transfer on superhydrophilic, superhydrophobic, and superbiphilic surfaces. *International Journal of Heat and Mass Transfer* **57**, 733–741 (2013).
12. Rahman, M. M., Ölçeroğlu, E. & McCarthy, M. Role of Wickability on the Critical Heat Flux of Structured Superhydrophilic Surfaces. *Langmuir* **30**, 11225–11234 (2014).
13. Song, Y. *et al.* Three-Tier Hierarchical Structures for Extreme Pool Boiling Heat Transfer Performance. *Advanced Materials* **34**, 2200899 (2022).
14. Theofanous, T. G., Tu, J. P., Dinh, A. T. & Dinh, T. N. The boiling crisis phenomenon: Part I: nucleation and nucleate boiling heat transfer. *Experimental Thermal and Fluid Science* **26**, 775–792 (2002).
15. Judd, R. L. On Nucleation Site Interaction. *Journal of Heat Transfer* **110**, 475–478 (1988).
16. Liu, J. *et al.* Bubble Coalescence During Pool Boiling with Different Surface Characteristics. *Heat Transfer Engineering* **45**, 360–380 (2024).

17. Hsu, Y. Y. On the Size Range of Active Nucleation Cavities on a Heating Surface. *Journal of Heat Transfer* **84**, 207–213 (1962).

18. Wang, C. H. & Dhir, V. K. On the gas entrapment and nucleation site density during pool boiling of saturated water. *Journal of Heat Transfer* **115**, 670–679 (1993).

19. Qi, Y., Klausner, J. F. & Mei, R. Role of surface structure in heterogeneous nucleation. *International Journal of Heat and Mass Transfer* **47**, 3097–3107 (2004).

20. Bankoff, S. G. Entrapment of gas in the spreading of a liquid over a rough surface. *AIChE Journal* **4**, 24–26 (1958).

21. Calka, A. & Judd, R. L. Some aspects of the interaction among nucleation sites during saturated nucleate boiling. *International Journal of Heat and Mass Transfer* **28**, 2331–2342 (1985).

22. Zhang, L. & Shoji, M. Nucleation site interaction in pool boiling on the artificial surface. *International Journal of Heat and Mass Transfer* **46**, 513–522 (2003).

23. Golobic, I., Petkovsek, J. & Kenning, D. B. R. Bubble growth and horizontal coalescence in saturated pool boiling on a titanium foil, investigated by high-speed IR thermography. *International Journal of Heat and Mass Transfer* **55**, 1385–1402 (2012).

24. Gjerkeš, H. & Golobič, I. Measurement of certain parameters influencing activity of nucleation sites in pool boiling. *Experimental Thermal and Fluid Science* **25**, 487–493 (2002).

25. Hutter, C. *et al.* Nucleation site interaction between artificial cavities during nucleate pool boiling on silicon with integrated micro-heater and temperature micro-sensors. *International Journal of Heat and Mass Transfer* **55**, 2769–2778 (2012).

26. Kangude, P. & Srivastava, A. On the Coupled Thermal and Hydrodynamic Interaction of Adjacently Located Vapor Bubbles on Highly Wetting Surfaces. *Langmuir* **38**, 13647–13658 (2022).

27. Kumar, A., Kangude, P. & Srivastava, A. Coupled bubble dynamics and interaction mechanisms of adjacently nucleated vapor bubbles under subcooled pool boiling regime. *Physics of Fluids* **35**, 087107 (2023).

28. Yu, C. *et al.* Manipulating Boiling Bubble Dynamics on Under-Liquid Superaerophobic Silicon Surfaces for High-Performance Phase-Change Cooling. *Adv Funct Materials* 2420677 (2025) doi:10.1002/adfm.202420677.

29. Park, H., Ahmadi, S. F., Foulkes, T. P. & Boreyko, J. B. Coalescence-Induced Jumping Bubbles during Pool Boiling. *Adv Funct Materials* **34**, 2312088 (2024).

30. Chen, T. & Chung, J. N. Heat–transfer effects of coalescence of bubbles from various site distributions. *Proc. R. Soc. Lond. A* **459**, 2497–2527 (2003).

31. Sultan, M. & Judd, R. L. Interaction of the Nucleation Phenomena at Adjacent Sites in Nucleate Boiling. *Journal of Heat Transfer* **105**, 3–9 (1983).

32. Kitron, A., Elperin, T. & Tamir, A. Stochastic modeling of boiling-site interaction. *Phys. Rev. A* **44**, 1237–1246 (1991).

33. Iwata, R. *et al.* How Coalescing Bubbles Depart from a Wall. *Langmuir* **38**, 4371–4377 (2022).

34. Dunlap, C., Li, C., Pandey, H., Le, N. & Hu, H. BubbleID: A deep learning framework for bubble interface dynamics analysis. *Journal of Applied Physics* **136**, 014902 (2024).

## Materials and Methods

### Microfabrication

The silicon test chips (2 cm × 2 cm, 380 μm thick) were microfabricated starting from a double-side polished silicon wafer. A 1 μm thick thermal $SiO_2$ layer was first grown on the wafer to electrically isolate the backside Pt microheater. The heater was then patterned over a 1 cm × 1 cm area by a standard photolithography followed by metal deposition and lift-off. Microcavities of controlled geometry were etched into the top boiling surface using deep reactive-ion etching. All cavities were positioned at approximately 1500 μm from the edge of the 1 cm × 1 cm heating area to ensure they reside within the thermally uniform region of the heater. More details can be found in **Figs. S2** and **S3** and **Supplementary Notes S2** and **S3**.

### Pool-boiling setup

We conducted all experiments with milli-Q water as the working fluid. Our setup consists of a glass chamber (100 mm × 100 mm × 200 mm) clamped between two PEEK fixtures. An immersion heater (100 W) heats the water to saturation temperature (100 °C), while two flexible rope heaters wrapped around the glass chamber exterior minimize parasitic heat losses to the environment. A reflux condenser mounted at the top continuously returns evaporated water to the pool, maintaining a constant liquid level (~1.5 L) and steady bulk temperature throughout each experiment.

### Cleaning and preconditioning protocol

Before loading each chip, we follow a rigorous cleaning protocol to ensure reproducible surface conditions. We rinse each chip sequentially in acetone, isopropanol, and deionized water, followed by 6–8 minutes of oxygen plasma treatment to remove organic residues and restore surface hydrophilicity. Contact angle measurements confirm that this protocol yields consistent wettability: advancing and receding contact angles measured before and after experiments remain within 15 ± 3° and 10 ± 2°, respectively, indicating that surface chemistry is preserved throughout the experimental campaign (**Fig. S8**). We rinse the glass chamber twice with deionized water, fill it with 1.5 L of fresh milli-Q water, and operate the immersion heater overnight (12–16 hours) to bring the pool to saturation temperature while driving off dissolved gases. After loading the cleaned chip, we apply a preconditioning heat flux of 50 $W \cdot cm^{-2}$ for 2–3 hours to remove residual gas trapped within the cavities and establish a reproducible nucleation baseline. Snapshots of representative chips imaged before and after experiments confirm that the cavity geometry and surface morphology remain intact throughout the experimental campaign (**Fig. S9**).

### Heat flux and temperature measurement

We powered the Pt thin-film heater using a DC power supply (0–600 V, 0–2 A). The same metal pattern also serves as an on-chip RTD, providing temperature measurement of the sample. We continuously acquired voltage and current data at 5 Hz using a custom electrical circuit with a voltage divider and shunt resistor (**Supplementary Note S4**). From the measured voltage $V$ and current $I$, we computed the real-time electrical resistance as $R = V/I$ and converted it to the temperature $T$ using a resistance–temperature curve calibrated prior to the boiling experiments (**Supplementary Note S5**). The input heat flux was obtained from $q''_{in} = V^2(R \cdot A_h)^{-1}$, where $A_h$ is the 1 cm × 1 cm heater footprint area.

### Boiling curve construction and heat flux correction

We constructed boiling curves by incrementally stepping the input heat flux and recording each data point once steady-state conditions were reached, holding at each step for at least 10 min before data acquisition. For each data point, we corrected the input heat flux $q''_{in}$ for lateral heat spreading and peripheral convection losses using a steady-state heat transfer model in COMSOL Multiphysics controlled through LiveLink for MATLAB, numerically iterating the boiling heat transfer coefficient until the simulated heater temperature matched the measured RTD value to within 0.1 °C. From the converged model, we extracted the corrected boiling heat flux $q''$ which we used to construct all boiling curves reported in this study.

**Data availability**
All data needed to evaluate the conclusions in the paper are present in the main article and the Supplementary Materials. Further data would be available upon reasonable request from the corresponding author.

**Acknowledgments**
We thank Cyrille Hilbert and Joffrey Pernolett for valuable discussions on microfabrication, and Laurent Chevalley for his assistance with the experimental setup fabrication.

**Funding**
This work was supported by the Swiss National Science Foundation under project grant number 219327.

**Competing Interests**
The authors declare no competing financial interests.

**Author contributions**

R.I. and Z.L. conceived the research and designed the study. R.I. performed the microfabrication of all test chips and, together with G.R., designed and built the pool boiling experimental setup including data acquisition procedures. R.I. and G.B. conducted the experiments. F.M., G.B., and Z.L. performed the numerical simulations. R.I. and G.B. carried out data analysis and post-processing. R.I. and Z.L. wrote the original manuscript. R.I., Z.L., and L.Z. contributed to figure conceptualization and preparation. Z.L. provided scientific guidance throughout the project. All authors reviewed and commented on the manuscript at various stages of preparation.

**Supplementary Materials**

**The PDF file includes:**
Figs. S1 to S9
Supplementary Notes S1 to S9
Legends for Videos S1 to S5
References

**Other Supplementary Material for this manuscript includes the following:**
Videos S1 to S5

Supplementary Information

# Scale-Separated Collective Bubble Nucleation and Departure

Rameez Iqbal *et al.*

*Corresponding author: zhengmao.lu@epfl.ch

**This PDF file includes:**

Figs. S1 to S9
Supplementary Notes S1 to S9
Legends for Videos S1 to S5
References

**Other Supplementary Materials for this manuscript include the following:**

Videos S1 to S5

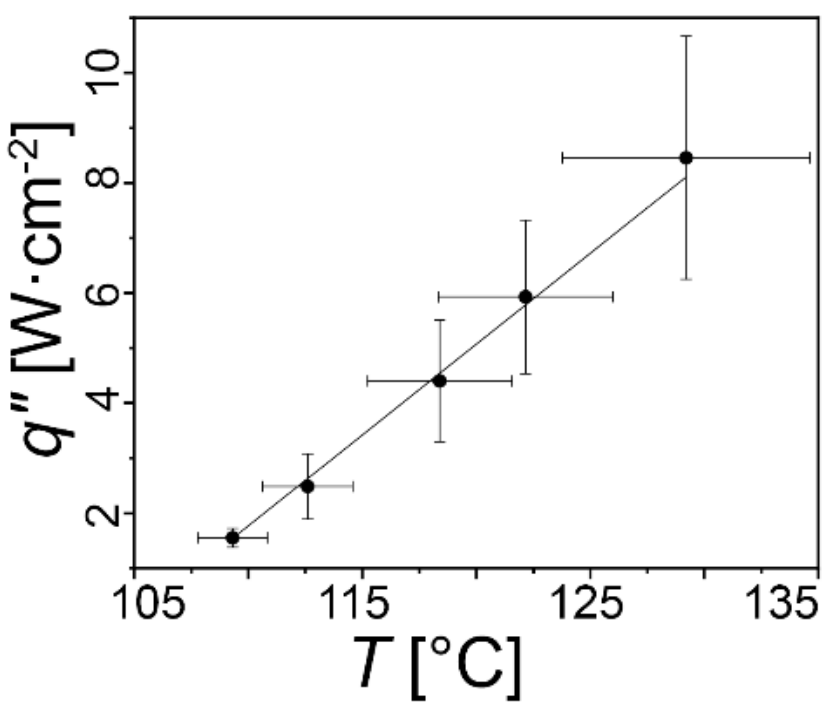

**Fig. S1 Thermal boundary layer thickness ($\delta_{th}$) from single-phase heat transfer.** Heat flux $q''$ as a function of sample temperature $T$ in the single-phase regime. Each data point represents the mean and standard deviation across more than 8 experimental runs and samples of different cavity patterns. The slope of the linear fit gives the single-phase heat transfer coefficient $h_{sp}$, from which the thermal boundary layer thickness is estimated as $\delta_{th} = k_w/h_{sp}$ (≈ 206 μm in our experiments), where $k_w$ is the thermal conductivity of water.

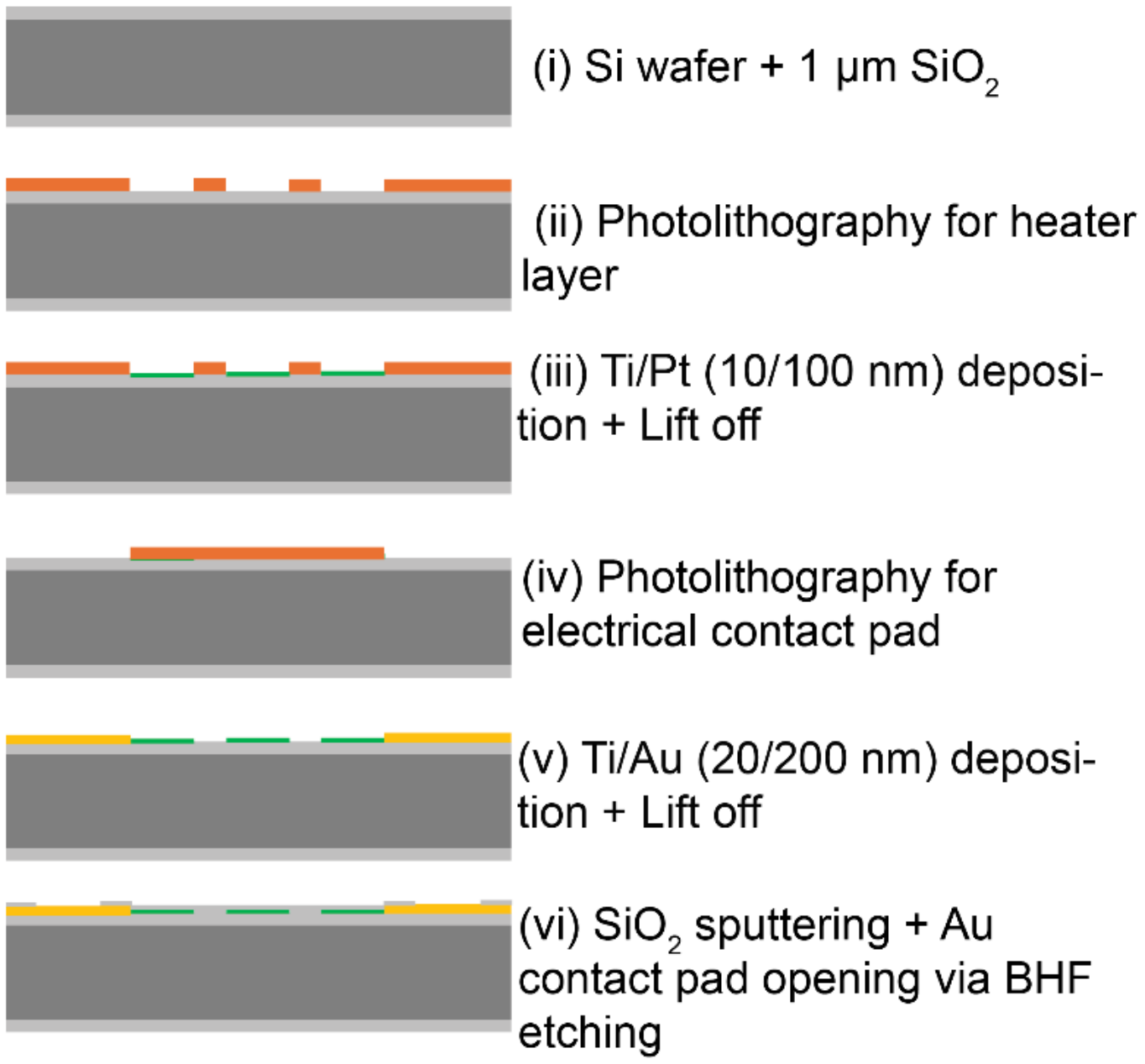

**Fig. S2 Process flow for microheater fabrication.** (i) 4-inch double-side polished Si wafer coated with 1 μm thermal $SiO_2$; (ii) photolithography for the heater layer; (iii) Ti/Pt (10/100 nm) deposition by e-beam evaporation and the subsequent lift-off; (iv) photolithography for the electrical contact pads; (v) Ti/Au (20/200 nm) deposition and the subsequent lift-off; and (vi) 30 nm $SiO_2$ passivation, followed by opening of Au contact pads via buffered HF (BHF) etching. The wafer is then annealed in forming gas (95% $N_2$ + 5% $H_2$, 400 °C, 60 min) to stabilize the Ti/Pt grain structure.

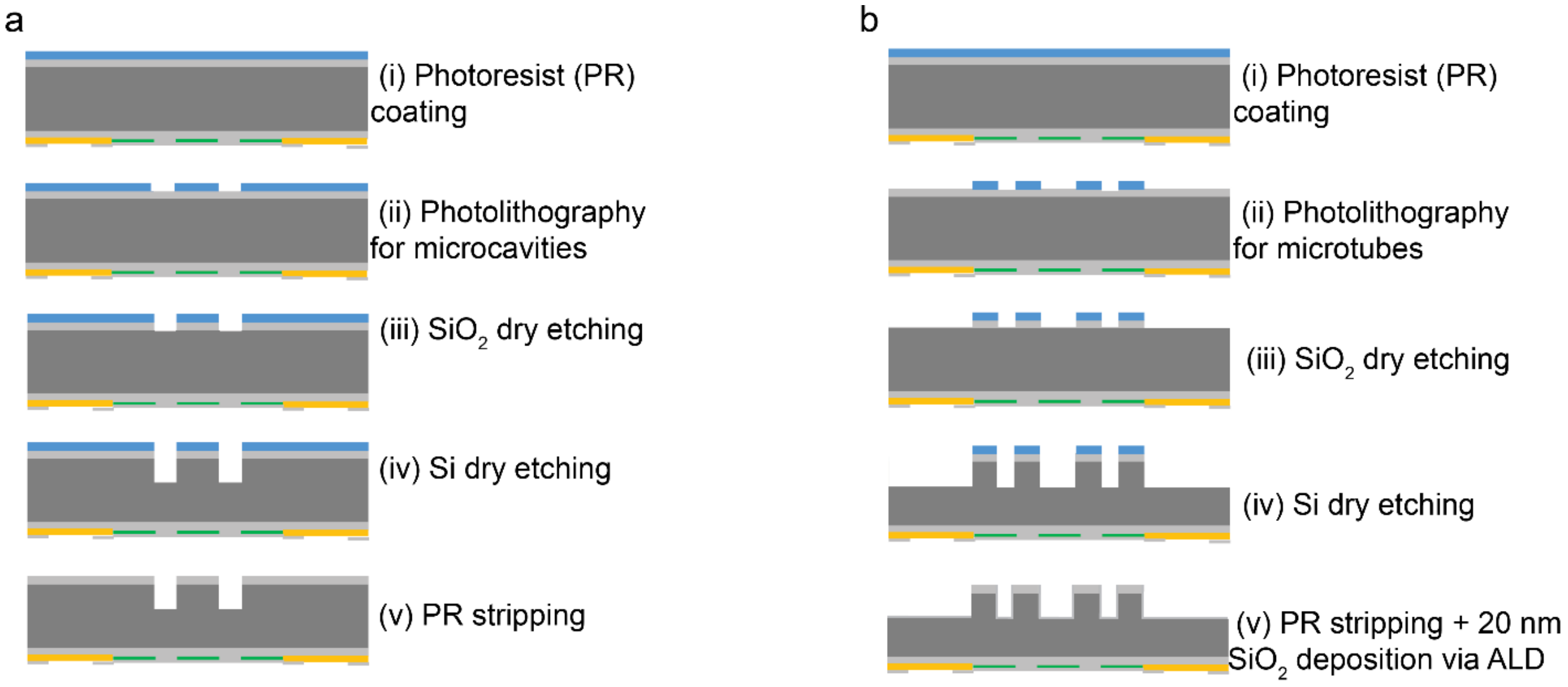

**Fig. S3 Fabrication process for (a) microcavities and (b) microtubes.** Steps (i)–(iv) are common to both: (i) photoresist (PR) coating, (ii) photolithography to define features, (iii) $SiO_2$ dry etching, and (iv) Si etching using deep reactive-ion etching (DRIE, Bosch process). (v) PR stripping for microcavities and for microtubes, PR stripping is followed by conformal deposition of 20 nm $SiO_2$ by atomic layer deposition (ALD).

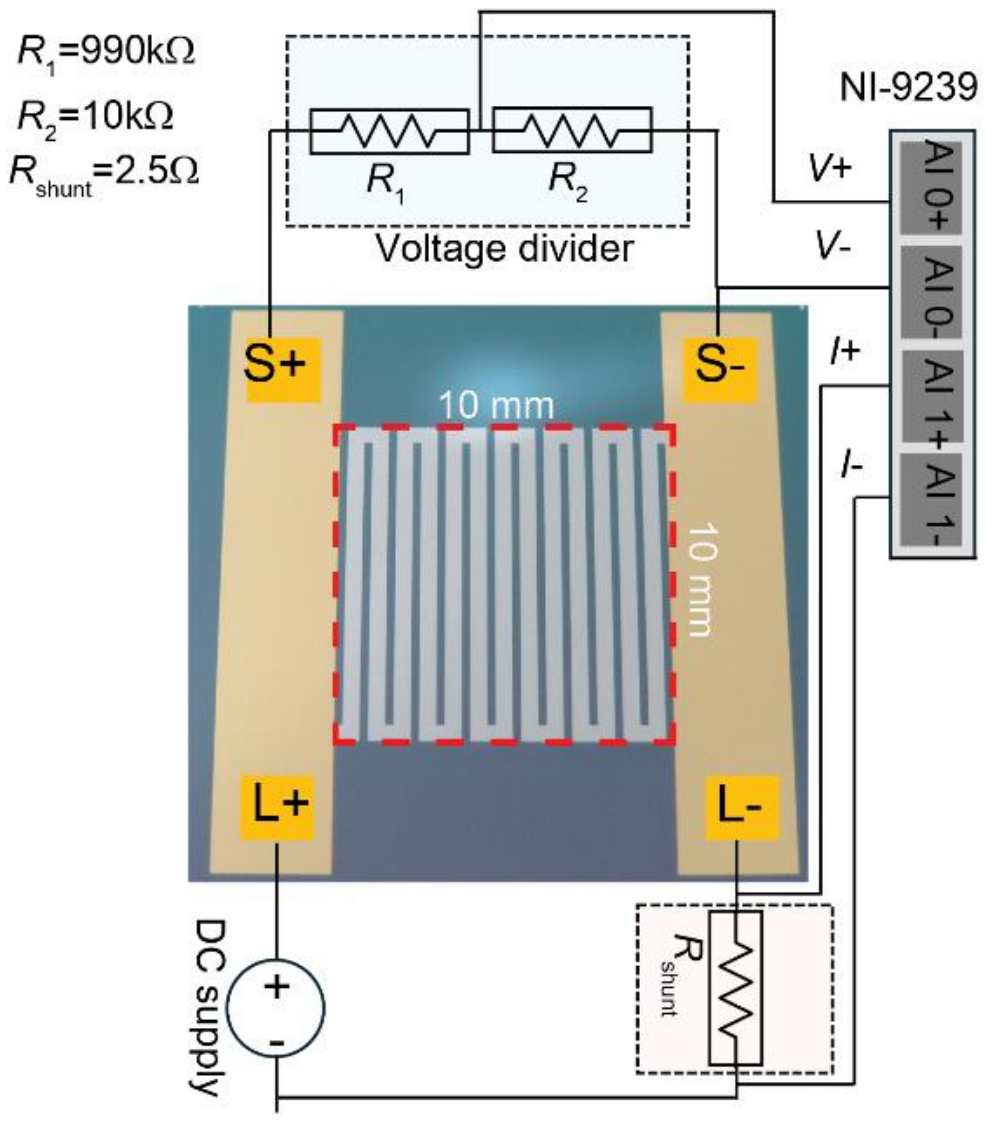

**Fig. S4 Measurement of input heat flux and sample temperature.** Electrical measurement circuit for simultaneous acquisition of heater voltage ($V$) and current ($I$) for the microheater (10 mm × 10 mm serpentine pattern; red dashed box). An ultra-precision high-voltage divider (Caddock USVD2, $R_1$ = 990 kΩ, $R_2$ = 10 kΩ) scales down heater voltage for the NI-9239 DAQ (±10 V range), and the current is measured from the voltage drop across an ultra-high-precision Z-foil shunt resistor (Vishay Precision Group, 2.5 Ω) in series with the microheater. The input heat flux is then determined from the electrical power dissipated in the heater divided by the heater area. We use 4-wire Kelvin sensing (L: load lines; S: sense lines) which eliminates the contact resistance at heater leads. The microheater simultaneously serves as an on-chip resistance temperature detector (RTD) for real-time temperature readout during boiling experiments.

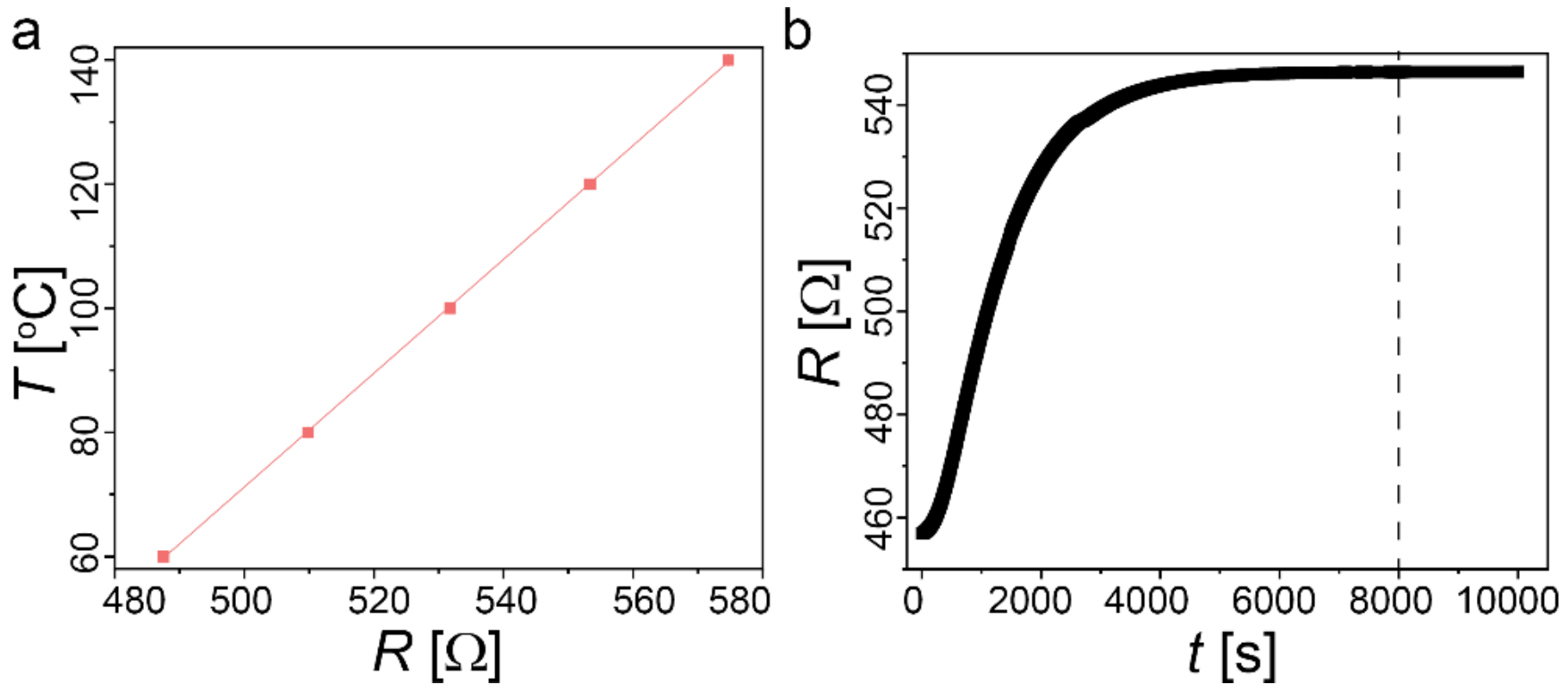

**Fig. S5 Temperature calibration of on-chip microheater.** (a) Sample-specific calibration (sample temperature *T* vs its electrical resistance *R*) of the on-chip microheater placed inside a temperature-controlled furnace. During the calibration, *T* was determined from a calibrated secondary-reference platinum resistance thermometer (Fluke 5609) placed adjacent to the sample. Temperature was computed on the ITS-90 scale using the probe's individual calibration-certificate coefficients (deviation function), rather than a nominal Pt100 (Callendar-Van Dusen) conversion. Each data point represents time-averaged resistance acquired over 600 s at each oven set point (60–140 °C) after 8000 s dwell time. The quadratic fit (solid red line) yields the calibration coefficients for actual temperature readout during experiments. (b) Temporal evolution of *R* at a fixed oven set point; the dashed line marks the 8000 s dwell time after which resistance was recorded, confirming that the chip reaches stable thermal equilibrium well before data acquisition.

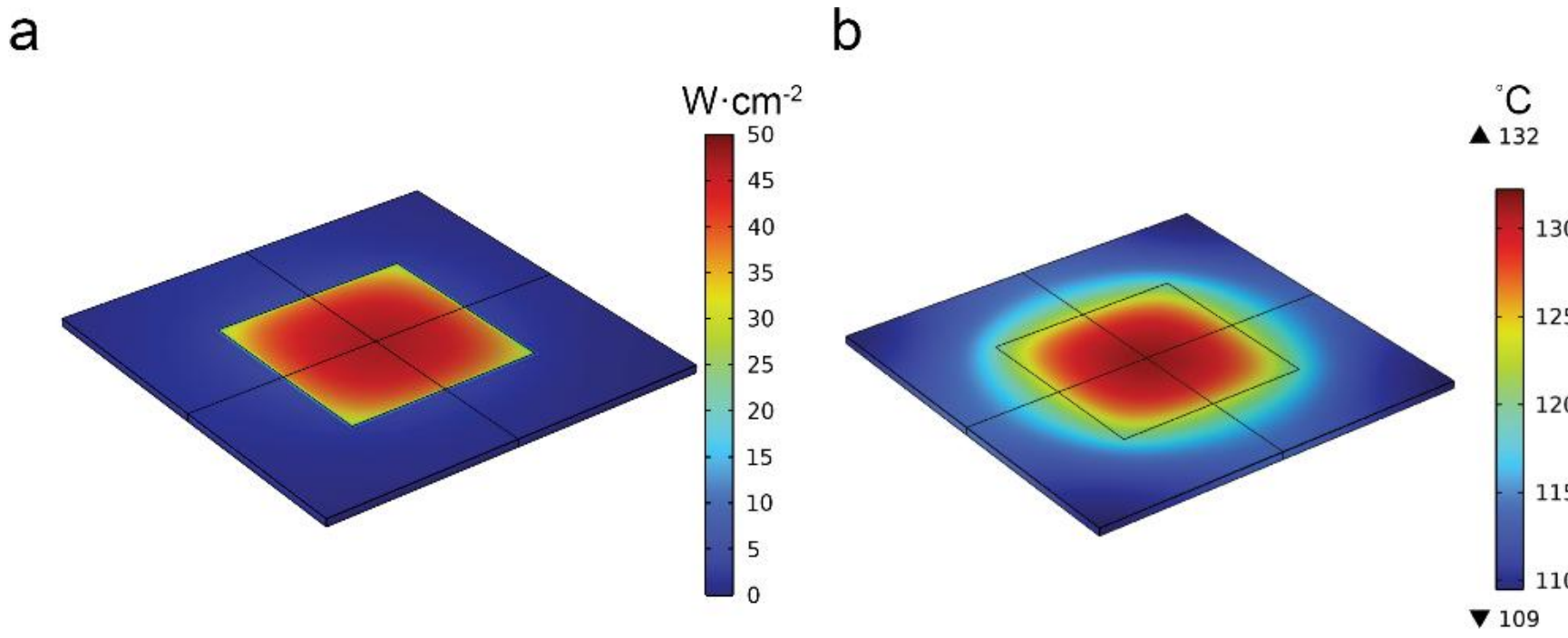

**Fig. S6 Determination of boiling heat flux.** For each experimental data point, a steady-state COMSOL Multiphysics heat conduction model iterates the boiling heat transfer coefficient over the 1 $cm^2$ central area via a bisection algorithm until the simulated backside heater temperature matches the measured RTD reading within 0.1 °C. Upon convergence, the model extracts the actual average boiling heat flux $q''$ over the 1 $cm^2$ central area. Representative result at $q''_{in}$ = 50 W·$cm^{-2}$: (a) heat flux distribution and (b) temperature distribution on the top boiling surface. See Supplementary Note S6 for full details.

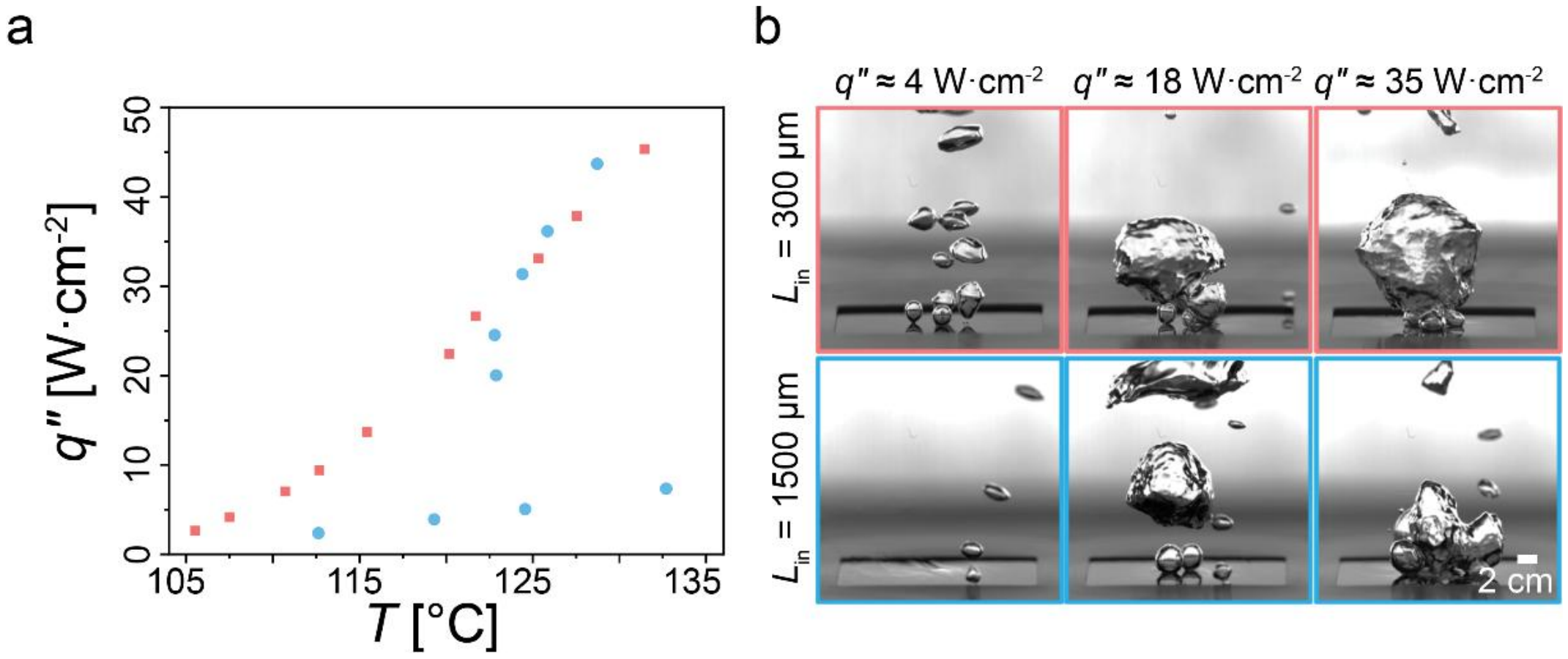

**Fig. S7 Nucleation and bubble dynamics for paired cavities with fixed inter-pair separation $L_D$ = 5000 µm.** (a) Boiling curves $q''$–$T$ for closely spaced pairs ($L_{in}$ = 300 µm, red squares) and widely spaced pairs ($L_{in}$ = 1500 µm, blue circles) during deactivation sweeps, analogous to **Fig. 4b**. (b) Representative side-view snapshots at low, intermediate, and high heat fluxes ($q'' \approx$ 4, 18 and 35 W·cm$^{-2}$) showing bubble dynamics for the two intra-pair spacings, analogous to **Figs. 1d** and **1f**.

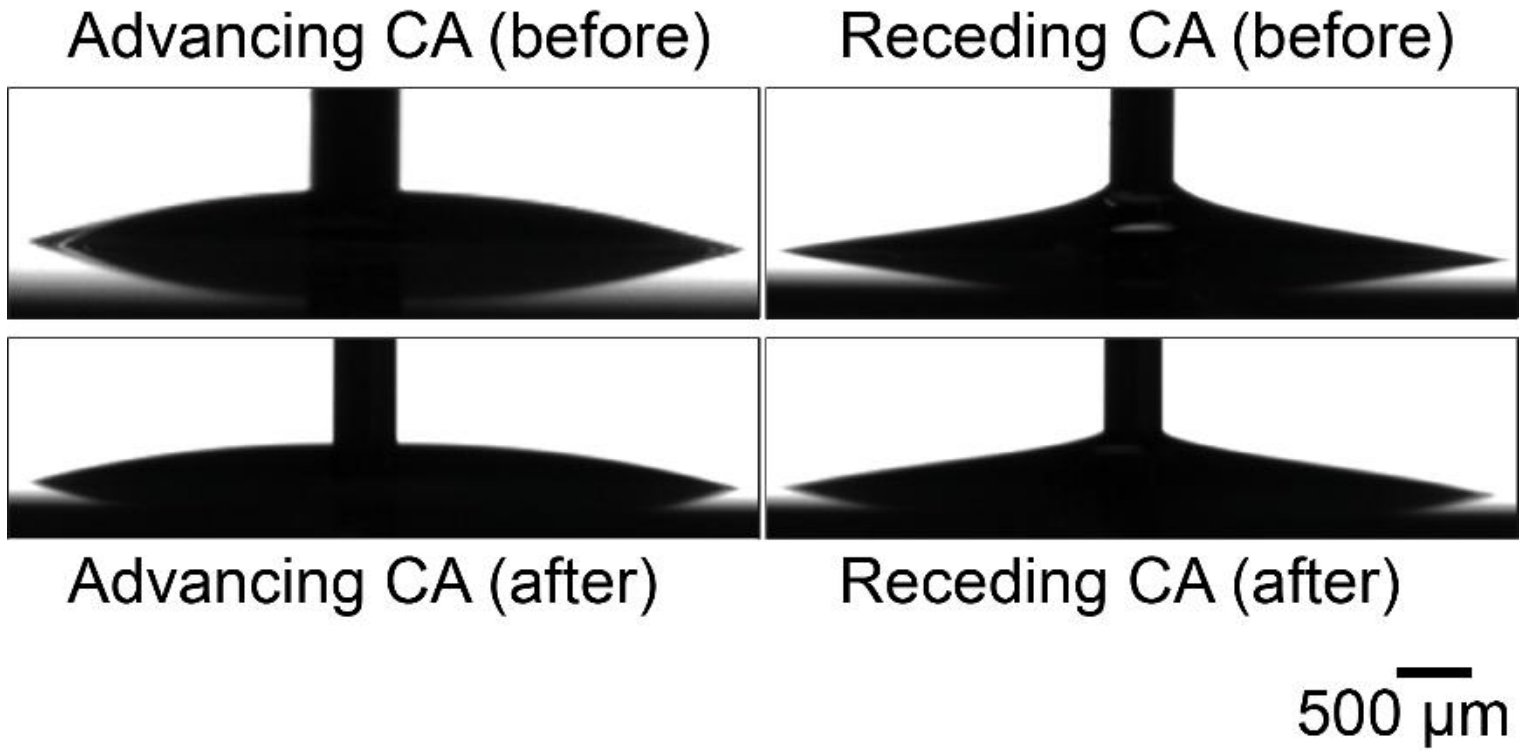

**Fig. S8 Wettability and contact angle (CA) characterization of the boiling surface.** Advancing contact angle (15 ± 3°) and receding contact angle (10 ± 2°) before and after pool boiling experiments show negligible change, confirming that surface wettability remains stable.

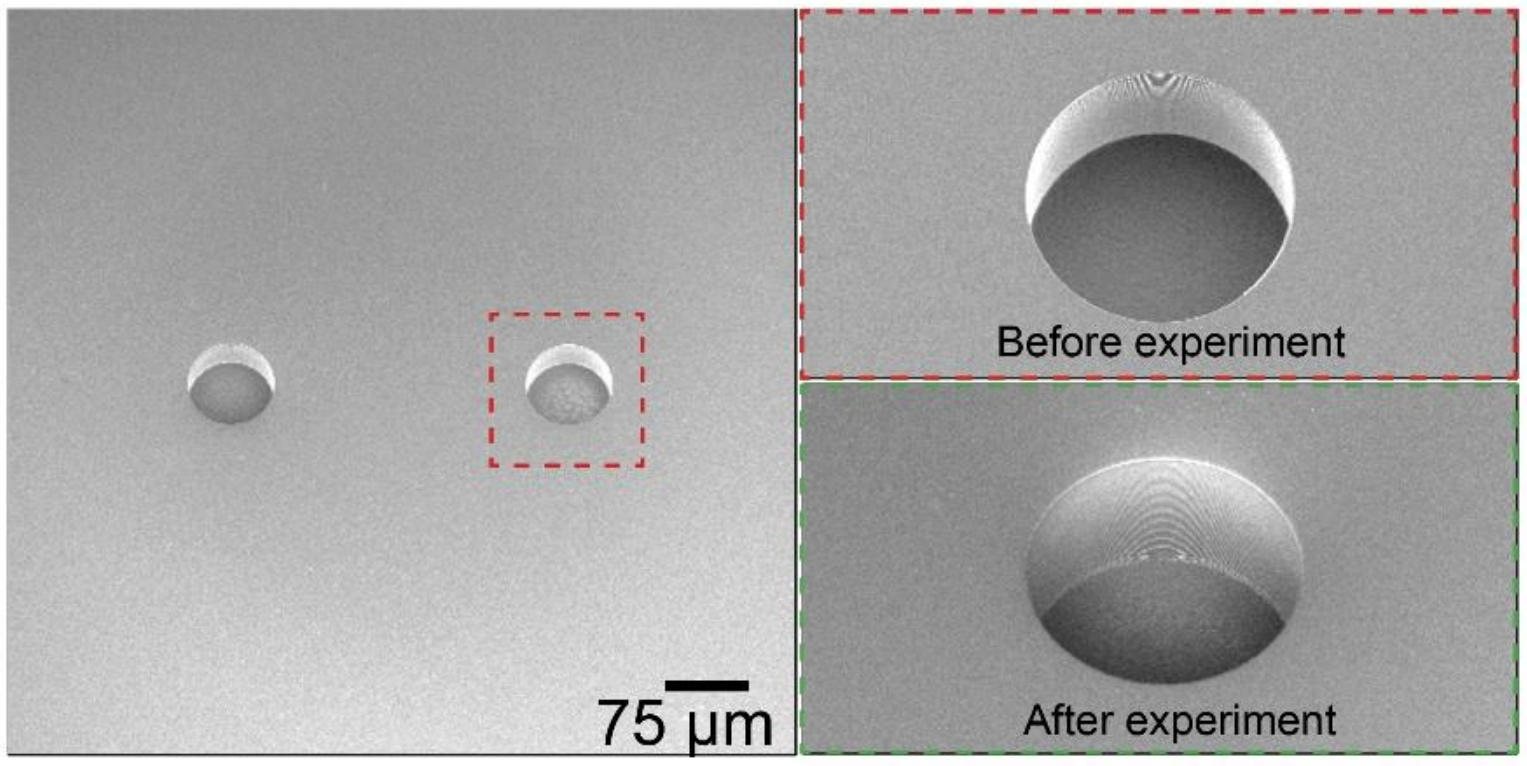

**Fig. S9 Surface morphology before and after pool boiling.** SEM images of the microfabricated boiling surface before (red dash) and after (green dash) pool boiling experiments. The unmagnified view shows the surface prior to any experiment. The magnified views confirm that the engineered microcavities retain their structural integrity with no significant degradation or contamination accumulation after repeated thermal cycling during pool boiling experiments.

**Supplementary Note**

## S1. Estimation of hydrodynamic boundary layer thickness

**Fig. S1** shows the single-phase heat flux as a function of sample temperature, measured before the onset of nucleate boiling. We averaged data across multiple samples (of different cavity patterns) and applied a linear fit to the resulting curve; the slope yields the single-phase heat transfer coefficient ($h_{sp}$), from which we estimated the effective thermal boundary layer thickness:

$$\delta_{\mathrm{th}} = \frac{k_{\mathrm{w}}}{h_{\mathrm{sp}}} \tag{S1}$$

where $k_w = 0.679\ \mathrm{W \cdot m^{-1} \cdot K^{-1}}$ is the thermal conductivity of saturated liquid water at 100 °C, giving $\delta_{th} = 206$ µm.

For the hydrodynamic boundary layer $\delta_H$, we applied the standard laminar natural convection scaling for a horizontal heated surface in quiescent liquid:

$$\delta_{\mathrm{H}} = \delta_{\mathrm{th}} \mathrm{Pr}^{1/3} \tag{S2}$$

where Pr is the Prandtl number evaluated for saturated liquid water at 100 °C, yielding $\delta_H \approx 248$ µm.

## S2. Fabrication of microheaters

We fabricated the microheaters on a 4-inch double-side polished silicon wafer (380 µm thick) coated with 1 µm of wet thermal oxide ($SiO_2$) on both sides. First, we spin-coated the wafer with 2 µm of AZ 10XT photoresist, exposed the heater patterns using a maskless aligner (MLA), and developed the patterns in AZ 400K: $H_2O$ (1:3.5) solvent. To form the resistive heating elements, we deposited a 10 nm Ti adhesion layer and a 100 nm Pt layer via e-beam evaporation, followed by lift-off in SVC-14 solvent remover. Next, we defined the electrical contact pads using a second photolithography and lift-off cycle, depositing 20 nm of Ti and 200 nm of Au via e-beam evaporation. We then sputtered a 30 nm $SiO_2$ passivation layer over the entire heater side and selectively removed it from the contact pad regions using buffered hydrofluoric acid. Finally, we annealed the wafer in a forming gas (95% $N_2$ + 5% $H_2$) environment at 400 °C for 60 min. This annealing step stabilizes the grain structure of the thin films, ensuring that the electrical resistance of the heating layer remains stable and equilibrated when subjected to high-current conditions during the boiling experiments.

## S3. Fabrication of microcavities and microtubes

After fabricating the microheater, we protected the heater side of the wafer by spin-coating AZ 10XT resist and overbaking it at 150 °C. On the opposite side, we spin-coated ECI-3027 photo resist, exposed microcavities or microtubes using the MLA, performed a post-exposure bake at 100 °C for 1 min, and developed the resist in AZ 726 MIF solvent. We then dry-etched the 1 µm $SiO_2$ layer using reactive ion etching with a $C_4F_8/H_2/He$ gas mixture and subsequently etched the underlying silicon by DRIE (Bosch process) using $SF_6$ and $C_4F_8$ gases, followed by complete photoresist removal. The resulting cavities had a diameter of 75 µm and a depth of 50 ± 5 µm. For the cavity chips, the entire surface except the cavity sidewalls remained covered with the original 1 µm $SiO_2$. For the microtube chips, the etching of bulk silicon removed most of this oxide, leaving $SiO_2$ only on the top surface of each microtube; to provide a $SiO_2$-terminated surface comparable to the cavity chips, we deposited an additional 20 nm $SiO_2$ conformally via atomic layer deposition (ALD).

## S4. Measurement of input heat flux and sample temperature

To determine the input heat flux ($q''_{in}$) and sample temperature ($T$) during the experiments, we measured the voltage and current across the backside Pt heater. Electrical power was supplied by a programmable DC power supply (Keysight N5752A), while all measurements were performed with a high-resolution 24-bit data acquisition module (NI-9239, $\pm$10 V range in a cDAQ-9171 chassis).

We used a four-wire (Kelvin) configuration at the heater: separate force lines carry the heating current, and separate sense lines, which draw negligible current, read the heater voltage, eliminating the contribution of lead and contact resistances. Because the DAQ module has a voltage limit of ±10 V, the heater voltage on the sense lines was stepped down by an ultra-precision voltage divider (Caddock USVD2, $R_1$ = 990 kΩ, $R_2$ = 10 kΩ) and recovered by multiplying the measured divider voltage by the divider ratio. The current was obtained from the voltage drop across an ultra-high precision shunt resistor (Vishay Precision Z-Foil, 2.5Ω) placed in series with the heater in the supply return path, giving $I = V_{shunt}/R_{shunt}$. We chose $R_{shunt}$ = 2.5 Ω to balance signal-to-noise ratio against shunt self-heating: a lower resistance would make the measurement overly sensitive to DAQ offset noise, while a higher resistance would cause excessive Joule heating in the shunt.

Using the measured $V$ and $I$, we continuously monitored the instantaneous heater resistance $R = V/I$, which is then converted to the sample temperature ($T$) using the chip-specific RTD calibration (Supplementary Note S5). The raw input heat flux $q''_{in}$ is obtained by dividing the total heating power by the heater footprint area (1 cm$^2$). As detailed in **Note S6**, we then converted $q''_{in}$ into the actual boiling heat flux $q''$ using numerical simulations.

## S5. RTD calibration for temperature measurement

We used the platinum (Pt) microheater on the backside of the chip as a resistance temperature detector (RTD) to determine the chip temperature during pool boiling experiments. Before each experiment, we placed the test chip in a temperature-controlled oven together with a calibrated secondary-reference probe (Fluke 5609) positioned in proximity to the test chip. We determined the heater chip and reference probe resistance at several oven set points between 60 °C and 140 °C using a four-wire measurement with a data acquisition RTD module (NI-9226 on a cDAQ-9171), allowing the system to equilibrate at each temperature for approximately 8000 s before recording data for 600 s. We derived the corresponding reference temperature from the recorded reference-probe resistance using the probe's individual ITS-90 calibration certificate coefficients (*1*), rather than a Callendar-Van Dusen Pt100 conversion. We then plotted the mean heater resistance $R$ against the ITS-90-derived reference temperature $T$ and fitted the data with a quadratic model (**Fig. S5**): $T = aR^2 + bR + c$. During boiling experiments, we continuously measured the heater resistance $R$ in real time and applied the calibration coefficients ($a$, $b$, and $c$) to recover the chip temperature ($T$).

## S6. Determination of actual boiling heat flux

The silicon test chips have a 2 cm × 2 cm area while the backside platinum microheater was patterned over the central 1 cm × 1 cm area. Because the heater covers only one quarter of the chip area, a fraction of the Joule heating power spreads laterally through the 380-μm-thick Si substrate and escapes via natural convection from the unheated 3 cm$^2$ peripheral top surface rather than reaching the central 1 cm$^2$ boiling area. To account for this lateral spreading, we determined every data point using a steady-state heat conduction model built in COMSOL Multiphysics and controlled via LiveLink for MATLAB. We applied the measured input heat flux ($q''_{in}$) at the bottom boundary of the 1 $cm^2$ heater area, imposed adiabatic boundary conditions on the lateral chip edges (in contact with an insulating PEEK holder), and prescribed a temperature-dependent natural convection heat transfer coefficient over the unheated 3 $cm^2$ peripheral area, obtained from single phase heat transfer data before onset of boiling. We treated the boiling heat transfer coefficient over the central 1 $cm^2$ area as an unknown parameter, which we determined using a bisection algorithm: for each experimental data point, the solver recalculated the thermal field iteratively until the simulated backside heater temperature converged to match our experimentally measured RTD temperature within 0.1 °C. For single-phase data prior to boiling onset, we instead iterated a uniform heat transfer coefficient across the entire top surface. Upon convergence, the model yielded the actual boiling heat flux $q''$ over the 1 $cm^2$ central area, which we subsequently used to construct boiling curves reported in this study.

## S7. Two-dimensional flow-field simulation

Two-dimensional stationary simulations of liquid flow around neighboring surface bubbles (**Fig. 3c**) were performed in COMSOL Multiphysics (Laminar Flow module) to provide a mechanistic illustration of the near-wall hydrodynamic contrast between closely and widely spaced bubble configurations.

The computational domain is a 2D rectangle, 4 mm in the streamwise direction and $5\delta_H$ in the wall-normal direction. Two spherical-cap bubble obstacles are placed at the bottom wall, centered at $x = \pm L_{in}/2$, and subtracted from the fluid domain. Each bubble cap, pinned at the cavity rim, is represented as a circular arc with radius $r_{bub} = D_{cav}/(2\sin\theta)$, where $D_{cav}$ = 75 μm is the cavity diameter and $\theta$ = 15° is the prescribed contact angle. Two configurations were simulated: $L_{in}$ = 300 μm and $L_{in}$ = 1500 μm.

The steady-state incompressible Navier–Stokes equations were solved. Fluid properties correspond to saturated liquid water at 100 °C, using COMSOL's built-in temperature-dependent correlations for dynamic viscosity and density. The inlet (left) boundary is prescribed with a Kármán–Pohlhausen quartic velocity profile representing the hydrodynamic boundary layer:

$$\frac{u}{U_\infty} = \begin{cases} 2\eta - 2\eta^3 + \eta^4 & \text{for } \eta < 1 \\ 1 & \text{for } \eta \geq 1 \end{cases} \tag{S3}$$

where $\eta$ is the distance from the wall normalized to $\delta_H$. This profile approximates the Blasius flat-plate boundary-layer profile. The free-stream velocity $U_\infty$ = 0.038 m/s is estimated from the characteristic natural-convection velocity scale $U_\infty \sim (g\beta\Delta T L_{sample})^{1/2}$, appropriate for the experimental configuration (the entire sample size $L_{sample} \approx$ 2 cm, wall superheat $\Delta T \approx$ 10°C, and the thermal expansion coefficient $\beta$ is evaluated for saturated liquid water at 100°C). Because the simulation results are reported as the normalized velocity $u/U_\infty$, the normalized flow pattern is insensitive to the precise value of $U_\infty$.

A zero-pressure condition is applied at the right (outlet) boundary. The top boundary uses a symmetry condition. The bottom wall (outside bubble footprints) is no-slip. Bubble surfaces are assigned a free-slip boundary condition, approximating the negligible tangential viscous stress at a vapor–liquid interface. Mesh refinement studies were performed to confirm that the reported normalized velocity fields are insensitive to further refinement.

These simulations illustrate the qualitative modification of the near-wall velocity field by bubble obstacles at two different spacings. They do not account for bubble growth and departure dynamics or three-dimensional effects. The steady-state, isothermal formulation is intended to isolate the hydrodynamic field contrast that motivates the shielding mechanism discussed in the main text.

### S8. Measurement Uncertainty

Both the sample temperature $T$ and input heat flux $q''_{in}$ are derived from the same voltage ($V$) and current ($I$) signals (**Fig. S4**). Hardware specifications give fixed relative errors $\delta V/V$ = 0.0343% and $\delta I/I$ = 0.0422%. The heater resistance $R = V/I$ carries the combined relative uncertainty:

$$\frac{\delta R}{R} = \sqrt{\left[\left(\frac{\delta V}{V}\right)^2 + \left(\frac{\delta I}{I}\right)^2\right]} \tag{S4}$$

The temperature uncertainty follows by differentiating the calibration $T = aR^2 + bR + c$ (**Supplementary Note 5**):

$$\delta T = |2aR + b|R \cdot \left(\frac{\delta R}{R}\right) \tag{S5}$$

For all our samples, Eq. S5 results in $\delta T < 0.3$ °C.

As to the input heat flux $q''_{in} = V \cdot I/A_h$, where $A_h$ is the 1 $cm^2$ heater footprint area, the same voltage and current errors combine with the geometric uncertainty

$$\frac{\delta q''_{\mathrm{in}}}{q''_{\mathrm{in}}} = \sqrt{\left[\left(\frac{\delta V}{V}\right)^2 + \left(\frac{\delta I}{I}\right)^2 + \left(\frac{\delta A_{\mathrm{h}}}{A_{\mathrm{h}}}\right)^2\right]} \tag{S6}$$

With $\delta A_h/A_h \approx 0.1\%$ for the lithographically patterned area, the overall relative uncertainty in the input heat flux is 0.11%, dominated by the geometric fabrication tolerance rather than electrical acquisition noise.

### S9. Bubble Departure Diameter Extraction using BubbleID

Bubble departure diameters $D_b$ reported in **Fig. 5** were extracted using BubbleID, an automated bubble detection and tracking framework. The model classifies each detected bubble into one of three states per frame: attached to the surface, detached and rising, or unknown. To ensure reliable segmentation under our specific imaging conditions including the optical contrast, backlighting configuration, and bubble morphologies, the model was fine-tuned on a manually annotated subset of high-speed images spanning a range of heat fluxes and bubble sizes. After fine-tuning, the model was applied to the full image dataset, and individual bubble instances were tracked across consecutive frames using the OC-SORT visual tracking algorithm.

For each bubble instance, the segmentation mask was used to calculate the 2D projected area $A$ and the perimeter $p$ of the bubble contour. To account for slight shape deformations prior to departure, we computed $D_b$ as an arithmetic mean of the area-equivalent and perimeter-equivalent diameters:

$$D_b = \frac{1}{2}\left(2\sqrt{A/\pi} + p/\pi\right) \tag{S7}$$

In our analysis, we retained only bubbles that completed a fully resolved departure cycle from attachment through detachment, prior to the onset of inter-pair coalescence.

**Videos S1-S5.**

**Video S1.** *Effect of intra-pair spacing on nucleation at low heat flux.*

High-speed visualization of pool boiling on surfaces with intra-pair spacing $L_{in}$ = 300 and 1500 μm at $q'' \approx 4$ W·cm⁻² during decreasing heat flux sweep. Surfaces with $L_{in}$ comparable to $\delta_H$ (≈ 248 μm) show collective, sustained nucleation while surfaces with $L_{in} \approx 6\delta_H$ exhibit "*no nucleation*" at the engineered sites due to the absence of hydrodynamic shielding. The videos were recorded at 4000 frames per second and played at 10 frames per second.

**Video S2.** *Effect of intra-pair spacing on nucleation at intermediate heat flux.*

High-speed visualization of pool boiling on surfaces with intra-pair spacing $L_{in}$ = 300 and 1500 μm at $q'' \approx 18$ W·cm⁻². While $L_{in}$ = 300 μm maintains full site occupancy ($N$ = 8), bubble nucleation is confined to a subset of cavities ($N$ < 8) for $L_{in}$ = 1500 μm, referred to as "*partial nucleation*". The videos were recorded at 4000 frames per second and played at 10 frames per second.

**Video S3.** *Inter-pair-spacing-dependent vapor removal regimes at low heat flux.*

High-speed side-view visualization comparing bubble dynamics on surfaces with inter-pair spacing $L_D$ = 3000, 5000, and 7000 μm at $q'' \approx 2.5$ W·cm⁻² with $L_{in}$ fixed at 300 μm. Even at this low heat flux, inter-cluster coalescence has already emerged for $L_D$ = 3000 μm, placing it in the promotive regime, while bubble clusters for $L_D$ = 5000 and 7000 μm remain in isolated departure.

**Video S4.** *Inter-pair-spacing-dependent vapor removal regimes at intermediate heat flux.*

High-speed side-view visualization comparing bubble dynamics on surfaces with inter-pair spacing $L_D$ = 3000, 5000, and 7000 μm at $q'' \approx 10$ W·cm⁻² with $L_{in}$ fixed at 300 μm. As $D_b$ grows, the normalized inter-cluster spacing $L_D/D_b$ shrinks: promotive interactions have intensified at $L_D$ = 3000 μm, $L_D$ = 5000 μm has entered the promotive regime achieving the lowest surface temperature, while $L_D$ = 7000 μm continues in isolated departure as $D_b$ remains small relative to $L_D$.

**Video S5.** *Inter-pair-spacing-dependent vapor removal regimes at high heat flux.*

High-speed side-view visualization comparing bubble dynamics on surfaces with inter-pair spacing $L_D$ = 3000, 5000, and 7000 μm at $q'' \approx 35$ W·cm⁻² with $L_{in}$ fixed at 300 μm. At this high heat flux, $L_D$ = 3000 μm has transitioned into the excessive coalescence regime where a merged vapor mass spans multiple pairs and obstructs liquid replenishment, while $L_D$ = 7000 μm has grown into the promotive regime as $D_b$ increases, converging toward $L_D$ = 5000 μm which remains promotive, underscoring that $L_D/D_b$, not $L_D$ alone, governs the prevailing bubble interaction mode for vapor removal.

**References**

1. B. W. Mangum, "Guidelines for realizing the International Temperature Scale of 1990 (ITS-90)" (NBS TN 1265, National Bureau of Standards, Gaithersburg, MD, 1990); https://doi.org/10.6028/NIST.TN.1265.